\documentclass[journal,10pt]{IEEEtran}
\usepackage{cite}

\usepackage{graphicx}
\usepackage{subfigure}
\usepackage{url}
\usepackage{times}
\usepackage{setspace} 
\usepackage{amsmath}   
\usepackage{amsfonts,amsthm,amssymb}
\usepackage{mathtools}
\usepackage{nicefrac}
\usepackage[none]{hyphenat}
\usepackage{fontenc}
\usepackage{color}
\usepackage{hyperref}
\usepackage{url}
\usepackage[normalem]{ulem}
\usepackage{balance}
\usepackage{etoolbox}
\usepackage{algorithm}
\usepackage{algorithmic}
\usepackage{enumitem}
\usepackage{dblfloatfix}
\usepackage[normalem]{ulem}
\usepackage{verbatim} 
\usepackage{float}
\usepackage{placeins}
\usepackage{accents}
\usepackage{lipsum}

\newlength{\dhatheight}
\newcommand{\doublehat}[1]{%
    \settoheight{\dhatheight}{\ensuremath{\hat{#1}}}%
    \addtolength{\dhatheight}{-0.35ex}%
    \hat{\vphantom{\rule{1pt}{\dhatheight}}%
    \smash{\hat{#1}}}}

\DeclareMathOperator{\xp}{\mathbf{x}_p}
\DeclareMathOperator{\yDL}{\mathbf{y}_{DL}}
\DeclareMathOperator{\yDLs}{\hat{\mathbf{y}}_{DL}}
\DeclareMathOperator{\yDLss}{\doublehat{\mathbf{y}}_{DL}}
\DeclareMathOperator{\yUL}{\mathbf{y}_{UL}}
\DeclareMathOperator{\yULs}{\hat{\mathbf{y}}_{UL}}

\DeclareMathOperator{\ricJ}{\mathcal{J}}

\usepackage[dvipsnames]{xcolor}

\renewcommand{\baselinestretch}{0.93}
\newlist{steps}{enumerate}{1}
\setlist[steps, 1]{label = Step \arabic*:}
\begin{document}
\title{IEEE 802.11ad Based Joint Radar Communication Transceiver: Design, Prototype and Performance Analysis}

\author{Akanksha~Sneh, Soumya~Jain, V~Sri~Sindhu, Shobha~Sundar~Ram,~\IEEEmembership{Senior Member~IEEE}, and Sumit~Darak 
\thanks{All authors are with the Indraprastha Institute of Information Technology Delhi, New Delhi 110020 India. E-mail: \{soumya20325,sri20326,akankshas,sumit, shobha\}@iiitd.ac.in.}%
}

\maketitle

\begin{abstract}

Rapid beam alignment is required to support high gain millimeter wave (mmW) communication links between a base station (BS) and mobile users (MU). The standard IEEE 802.11ad protocol enables beam alignment at the BS and MU through a lengthy beam training procedure accomplished through additional packet overhead. However, this results in reduced latency and throughput.  Auxiliary radar functionality embedded within the communication protocol has been proposed in prior literature to enable rapid beam alignment of communication beams without the requirement of channel overheads. In this work, we propose a complete architectural framework of a joint radar-communication wireless transceiver wherein radar based detection of MU is realized to enable subsequent narrow beam communication. We provide a software prototype implementation with transceiver design details, signal models and signal processing algorithms. The prototype is experimentally evaluated with realistic simulations in free space and Rician propagation conditions and demonstrated to accelerate the beam alignment by a factor of four while reducing the overall bit error rate (BER) resulting in significant improvement in throughput with respect to standard 802.11ad. Likewise, the radar performance is found to be comparable to commonly used mmW radars. 
\end{abstract}

\section{Introduction}
An overarching objective of next generation intelligent transportation systems is to enable vehicle-to-everything (V2X) capabilities along key transport routes. This is to encourage sharing of road and vehicle information pertaining to environmental sensing for reducing road fatalities, and improving driving conditions eventually leading to autonomous driving \cite{hobert2015enhancements}. Currently, three modes of vehicular communications have been identiﬁed: \emph{dedicated short-range communication services (DSRC)} on IEEE 802.11p based wireless technology \cite{kenney2011dedicated}; \emph{device-to-device} (D2D) based V2X communications in long-term evolution (LTE) \cite{asadi2014survey}, and \emph{cellular LTE-V2X communications} \cite{wang2018cellular}. All three modes operate in sub-6 GHz spectrum; are restricted to tens of Megabits per second (Mbps) data rates and oﬀer latency of the order of several hundred milliseconds (ms). However, sharing of time-critical high deﬁnition three-dimensional environmental maps of congested road conditions between vehicles requires Gigabits per second (Gbps) data rates and ultra-low latency. 

Millimeter wave (mmW) unlicensed spectrum (above 24 GHz) oﬀers a viable solution for high bandwidth connected vehicles \cite{choi2016millimeter}. However, there are major challenges associated with the practical deployment of mmW transceivers. Due to the high propagation loss at these carrier frequencies, they are meant to operate in short-range line-of-sight (LOS) conditions with highly directional beams realized through beamforming. In high mobility environments, rapid beam training and management result in considerable overhead and signiﬁcant deterioration of latency \cite{liu2017millimeter}. Alternatively, auxiliary sensors such as GPS or standalone radars can aid in beam alignment of the communication systems \cite{aviles2016position}. However, the deployment of auxiliary sensors increases the cost and complexity in terms of synchronization and data processing as well as poses challenges in terms of interference. In this work, we propose to augment radar functionality within the communications transceiver to enable rapid beam alignment of the communication beams.

Joint radar communication (JRC) systems have been extensively explored over the last several decades over three broad paradigms. The first set of works considered the \emph{coexistence} of radar and communications systems on a common spectrum prompted by RF congestion issues \cite{li2017joint,martone2017spectrum,hassanien2016signaling,hessar2016spectrum}. Here, the main objective is to mitigate the mutual interference between the two systems with \cite{martone2017spectrum} or without cooperation \cite{hassanien2016signaling}. Examples include communication systems that operate as cognitive radio on federally allocated radar bands \cite{hessar2016spectrum}; or cognitive radar that operates on communication bands \cite{mishra2019cognitive}. In the second paradigm, \emph{passive radar} receivers are developed and deployed with communication transmitters serving as opportunistic illuminators for surveillance and remote sensing applications \cite{zeng2016wireless}. In the third paradigm, RF front ends - sharing spectrum and hardware - are \emph{co-designed} for both radar sensing and communication functionalities \cite{Hassanien2019dual,ram2022optimization}. Here the communication signals are embedded within a traditional radar framework \cite{sahin2017novel,jamil2008integrated} or the radar signals are embedded within the communication uplink or downlink signals \cite{liu2018toward}. In the other frameworks, either communication or the radar signal is treated as an interference signal or an uncooperative channel \cite{chiriyath2017radar} and hence the functionality of the two systems is not fully exploited as in the co-design approach. Our work focuses on co-design of an IEEE 802.11ad based dual-functional transceiver that is capable of joint radar remote sensing and communications. 

The IEEE 802.11ad standard specifies the media access control (MAC) and physical layer (PHY) protocols for implementing wireless local area networks through high data rate short range directional communication in the mmW spectrum \cite{noauthor_ieee_2016-1}. Preliminary studies have demonstrated the eﬀectiveness of JRC implemented upon IEEE 802.11ad for dynamic target tracking \cite{grossi2018opportunistic,kumari2019adaptive,duggal2020doppler}. In \cite{grossi2018opportunistic}, the authors proposed a method of using the IEEE 802.11ad beamforming training protocol for detecting radar targets. In \cite{kumari_ieee_2018}, the authors leveraged the high data rate (1.76GSa/s) and the Golay sequences within the channel estimation fields of the single carrier preamble in the IEEE 802.11ad packet to demonstrate radar ranging with fine range resolution and low sidelobes of a single point target. The authors in \cite{duggal2019micro, duggal2020doppler} proposed a modification to the preamble structure to enable Doppler resilient radar ranging of extended dynamic automotive targets. All of these works focused on developing radar signal processing algorithms without complete architectural details of the JRC transceivers and system-level performance analysis. Further, these works make assumptions on the radar waveform that would introduce significant limitations on the performance of the joint transceiver in realistic conditions. From a prototype perspective, there has been limited work in the design, prototype and experimental demonstrations of co-designed multi-functional transceivers \cite{ma2021spatial,kumari2021jcr70,pegoraro2021rapid}. In \cite{ma2021spatial}, a hardware prototype of a JRC system was demonstrated where radar and communications transmit waveforms were realized with index modulation via generalized spatial modulation. The transmitting antennas were divided into two sub-arrays, one for radar and the other for communication. In \cite{kumari2021jcr70}, a fully digital JRC architecture on hardware was demonstrated with a common transmitter for radar and communication, a monostatic radar receiver, and a separate communication receiver. A hardware prototype was also developed in \cite{pegoraro2021rapid}, where IEEE 802.11ay access points were retrofitted
for human detection and sensing in an indoor scenario. To the best of our knowledge, there is no software or hardware prototype for end-to-end JRC co-designed transceiver in the prior art. 

In this work, we propose a complete end-to-end architecture followed by a software implementation of a joint radar-communication framework based on IEEE 802.11ad for rapid beam alignment that is capable of half duplex communication and three-dimensional radar processing along range, azimuth and Doppler. The contributions of the paper can be summarized as follows:
\begin{enumerate}
    \item Standard IEEE 802.11ad packet consists of optional beam refinement fields allocated for beamforming training to locate the mobile user \cite{noauthor_ieee_2016-1}. The length of these fields is directly proportional to the number of beams supported by the transmitter. Compared to standard beam alignment for IEEE 802.11ad, we demonstrate that our proposed architecture accelerates the beam alignment by a factor of four by omitting the beam training fields.
    \item Standard IEEE 802.11ad framework utilizes a lengthy beam training procedure between the access point/base station (BS) \footnote{In this work, we use the terms base station and access point interchangeably} and the mobile user (MU) \footnote{In this work we use the terms mobile user and radar target interchangeably. Also, static clutter is differentiated from radar target.}. In our work, we implement three-dimensional (range-Doppler-Angle of Arrival) radar signal processing instead of the cumbersome beam training procedure. Through range processing, we detect point and extended targets in the radar field-of-view; through Doppler processing, we identify MU from static clutter; and through digital beamforming, we directly estimate the angle of arrival (AoA) of the MU.
    \item We provide a complete end-to-end prototype of the IEEE 802.11ad based JRC system realized using Matlab. This includes the design of BS with communication transmitter and receiver architecture along with the radar signal processor. We provide an in-depth discussion on the relevant radar and communication signal models, design details and signal processing algorithms. We have made our source codes open for interested readers. 
    \item  The proposed software prototype is experimentally evaluated for simulated point and extended targets under free space and Rician channel conditions and its communication metrics (throughput and latency) are benchmarked against the standard IEEE 802.11ad protocol. The radar performance metrics are likewise benchmarked against frequency modulated continuous waveforms commonly used in mmW radars.
\end{enumerate}
The paper is organized as follows. In Section~\ref{sec:SysArchitecture}, we present a comprehensive overview of the system architecture of the proposed JRC transceiver followed by its signal modeling in Section.\ref{sec:signalmodel}. In  Section.\ref{sec:simsetup}, we present the simulation setup for evaluating the performance of the proposed prototype. In Section.\ref{sec:Results}, we present the resultant beam alignment timing realized by the proposed radar-enhanced IEEE 802.11ad and contrast it with the standard IEEE 802.1ad beam alignment procedure. Further, we present the communication link metrics and the radar performance metrics for different types of targets (point and extended) along different trajectories. We conclude the paper with a summary of the findings and a discussion on future works in Section.\ref{sec:Conclusion}.\\
\emph{Notation:} Vectors and matrices are indicated with boldface lower and upper case characters respectively, while variables are indicated with regular characters. Radar and communication transmitted signals are indicated by $\mathbf{x}$ and $\mathbf{y}$ respectively while the received signals that have undergone one-way and two-way propagation are shown with $\hat{\mathbf{y}}$ and $\doublehat{\mathbf{y}}$ respectively. Vector superscript $T$ and $\ast$ denotes transpose and complex conjugate respectively. We use the square braces, $[\cdot]$, to indicate discrete-time sequences, and the curly braces, $(\cdot)$, to indicate continuous time signals. The terms BS-TX and BS-RX denote the transmitter and receiver, respectively, at the BS while MU-TX and MU-RX denote the transmitter and receiver at the non-access point station or MU. The uni-directional communication between BS-TX and MU-RX is referred to as downlink (DL) while the communication between MU-TX and BS-RX is referred to as uplink (UL).
\section{System Architecture of the JRC Transceiver}
\label{sec:SysArchitecture}
Directional communication links are required to overcome the atmospheric attenuation associated with mmW propagation. Hence, BS and MU must each determine the best beams prior to $\yDL$ and $\yUL$ communication. The standard protocol utilizes in-packet training for selecting the optimum pair of pre-determined beams at the BS and MU \cite{tsang2011coding}. In our proposed work, the radar functionality within the BS is used for beam alignment instead to reduce the overall beam alignment time. 
In this section, we present an overview of both the standard and the proposed JRC architectures and emphasize the modifications introduced in the former to realize the latter.
\subsection{Standard 802.11ad Transceiver Architecture}
\begin{figure}[!b]
    \centering
    \includegraphics[scale=0.75]{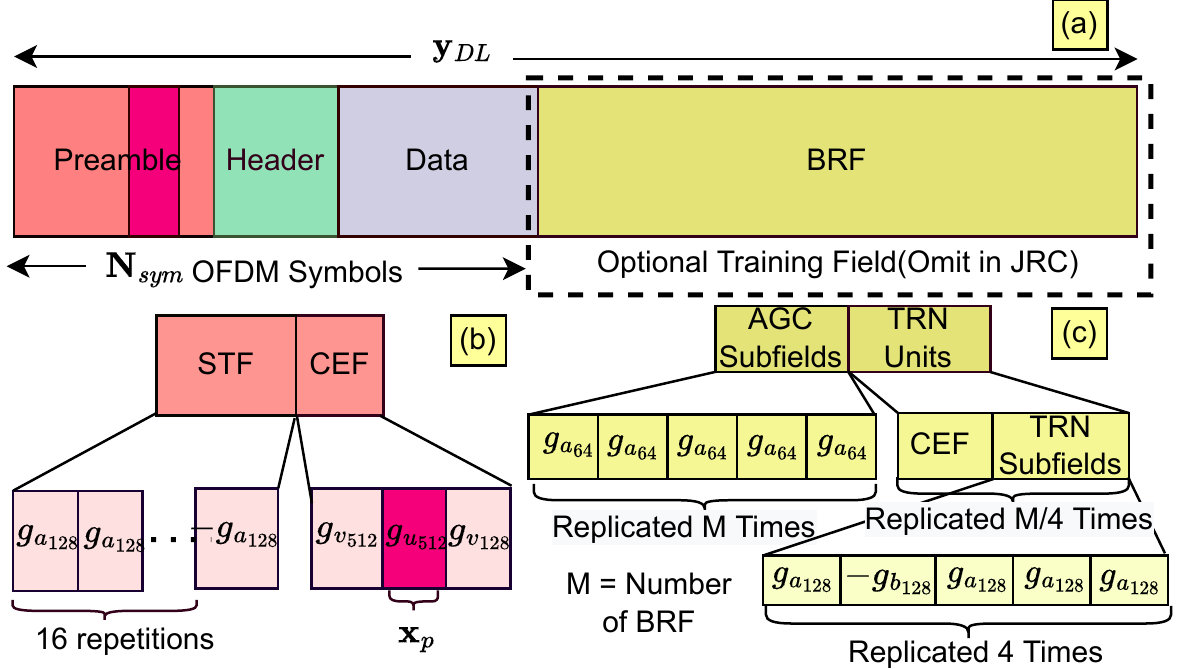}
    \vspace{-4mm}
    \caption{(a) IEEE 802.11ad JRC transmit packet structure; Expanded (b) preamble structure and (c) BRF structure.}
    \label{fig:jrcwaveform}
    \vspace{-2mm}
\end{figure}

We begin by providing a brief description of the PHY layer of the IEEE 802.11ad packet structure \cite{noauthor_ieee_2016-1}. 
The IEEE standard enables data and header transmission in three different modes: control, single carrier (SC), both at 1.76GHz sampling frequency, and orthogonal frequency division multiplexing (OFDM) mode at 2.64GHz sampling frequency \cite{noauthor_ieee_2016-1}. We model the data in our entire analysis in OFDM mode to enable high communication throughput and low latency.  As shown in Fig.\ref{fig:jrcwaveform}(a), a packet consists of preamble, header, data, and beam refinement protocol (BRF). 
The preamble has Golay sequences that exhibit perfect auto-correlation properties (with zero sidelobes) which make them suitable for radar remote sensing and channel estimation for communication \cite{kumari_ieee_2018}. The preamble is followed by a header which provides necessary control information to decode and demodulate the received data. The header is followed by data after which BRF fields with Golay sequences are provided for enabling beam training between BS and MU. The number of BRF fields has to be in multiples of 4 up to a maximum of 64 fields. 

\begin{figure*}[htbp]
\vspace{-4mm}
    \centering
    \includegraphics[scale=0.53]{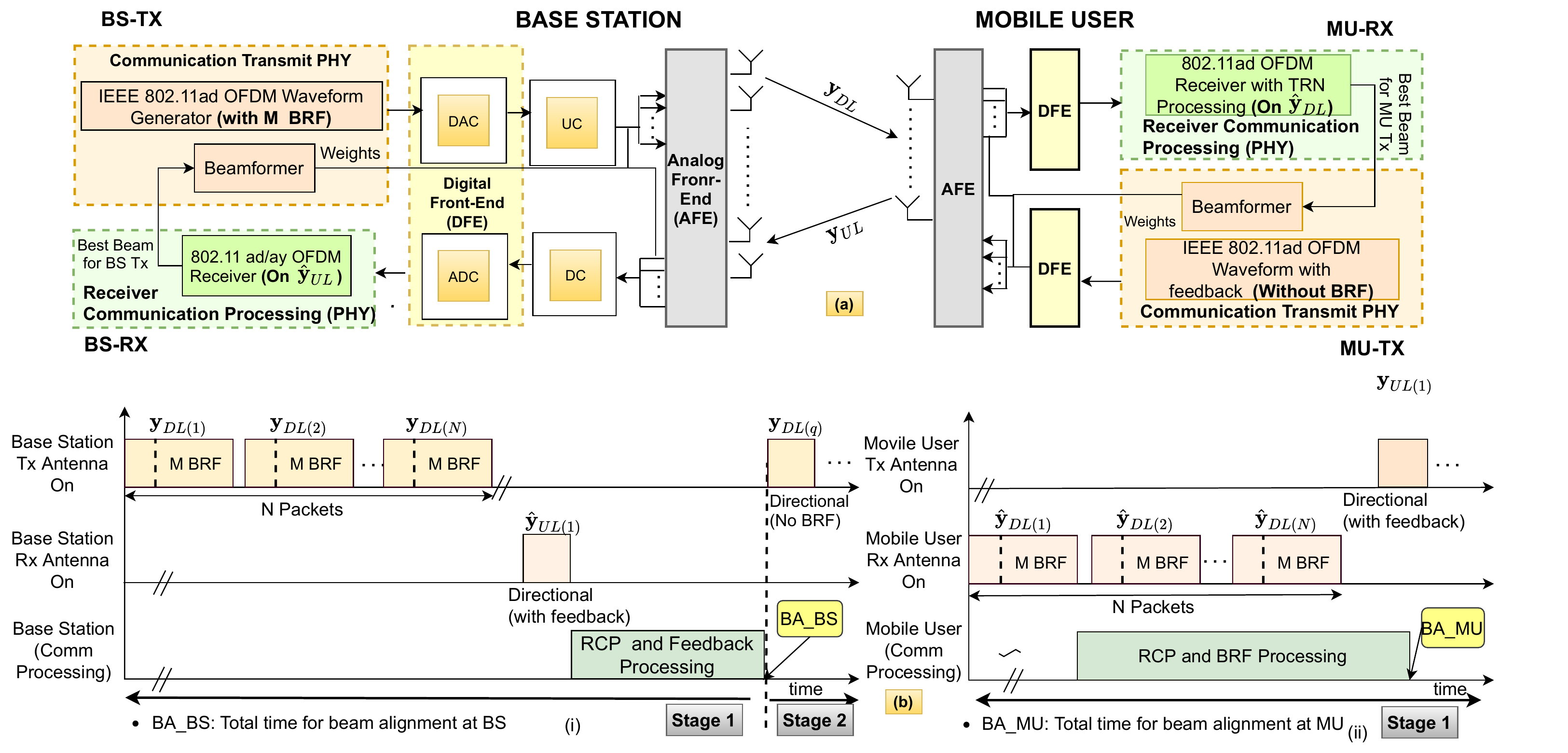}
    \vspace{-2mm}
    \caption{(a) Standard 802.11ad system architecture and (b) corresponding timing diagram for \emph{Stage-1} beam alignment based on beam training procedure.}
    \label{fig:std_arc}
    \vspace{-2mm}
\end{figure*}

The wireless transceiver architecture for supporting standard 802.11ad communication is presented in Fig.\ref{fig:std_arc}(a). 
The 802.11ad packets ($\mathbf{y}_{DL}$) are generated in the BS-TX and subsequently passed through the digital front end (DFE) consisting of a digital-to-analog converter (DAC).
Subsequently, the analog signal is modulated with an RF carrier in the analog front end (AFE), amplified and transmitted through a transmit phased antenna array. During the beam alignment phase, which we will henceforth refer to as \emph{Stage 1}, the preamble, header and data of the $\yDL$ packet are transmitted through a quasi-omnidirectional beam. $M$ additional BRF fields are appended to the $\yDL$ packet and transmitted along $M$ distinct pre-determined beams supported by the BS-TX phased array. Suitable antenna weight vectors (AWV) are applied at the BS-TX phased array to steer each BRF along the corresponding beam direction. The received signal at the phased array on the MU-RX, $\hat{\mathbf{y}}_{DL}$, is down-converted, digitized and processed to obtain the data. Based on the number of antennas in the MU-RX, $N$ total beams may be supported through analog beamforming at the MU. Based on \cite{tsang2011coding}, the BS will repeat the transmission of the $\yDL$ packet $N$ times and each packet will be received sequentially by different beams at MU as shown in the timing diagram in Fig.\ref{fig:std_arc}(b). To summarize, $M$ training fields in the $\yDLs$ packet are received through $N$ beams of the MU-RX leading to an overall complexity of $M \times N$. The best beam pair for the BS-TX and MU-RX will be estimated after the $N$ packets are processed at the MU-RX. If we assume reciprocity in the propagation environment, then the beam alignment procedure between BS-TX and MU-RX for $\yDL$ will provide the necessary information to support the $\yUL$ communication between MU-TX and BS-RX (as the same antennas support both the TX and RX functionalities at both BS and MU in half-duplex mode). The total duration for the beam alignment is the duration of the $N$ DL packets each appended with M training fields, the propagation time between the BS and MU and the receiver communication processing (RCP) time at the MU. Once the beam alignment is completed and communicated to the BS, we enter \emph{Stage 2}, where the data packets are transmitted between the BS and MU along the best beam pair without the need for BRF. The beam alignment process is repeated again depending on the mobility of the MU and the corresponding signal-to-noise ratio (SNR) estimated at the BS and MU becomes poor.

\begin{algorithm*}
\small{
\caption{Beam alignment procedure for standard 802.11ad}\label{alg:std_ta_steps}
\begin{steps}[leftmargin=1.5cm, label=Step \arabic*:]
  \item BS-TX sends $n = 1,2 \cdots,N$ downlink packets, each with $M$ BRF fields. Each $m^{th}$ BRF field is directed along $m^{th}$ pre-determined BS beam.
  \item MU-RX receives $N$ packets along $n = 1, 2, \cdots, N$ pre-determined MU beams.
  \item The $N$ received packets are processed at MU-RX (RCP) sequentially to determine best beam pair $(\tilde{m}, \tilde{n})$.
  \item MU-TX communicates about the $\tilde{m}$ beam to BS through $\mathbf{y}_{UL}$ without BRF field over the $\tilde{n}^{th}$ beam.
  \item The BS-RX receives and processes $\hat{\mathbf{y}}_{UL}$ and BS learns about its best beam ,i.e. $\tilde{m}$ for subsequent \emph{Stage-2}. This completes \emph{Stage -1} for the standard architecture.
  \end{steps}}
  \vspace{-1mm}
\end{algorithm*}

The steps for beam alignment in \emph{Stage-1} are summarized in Algorithm\ref{alg:std_ta_steps}.
\subsection{Proposed JRC Based 802.11ad Transceiver}
\begin{figure*}[htbp]
    \centering
    \includegraphics[scale=0.53]{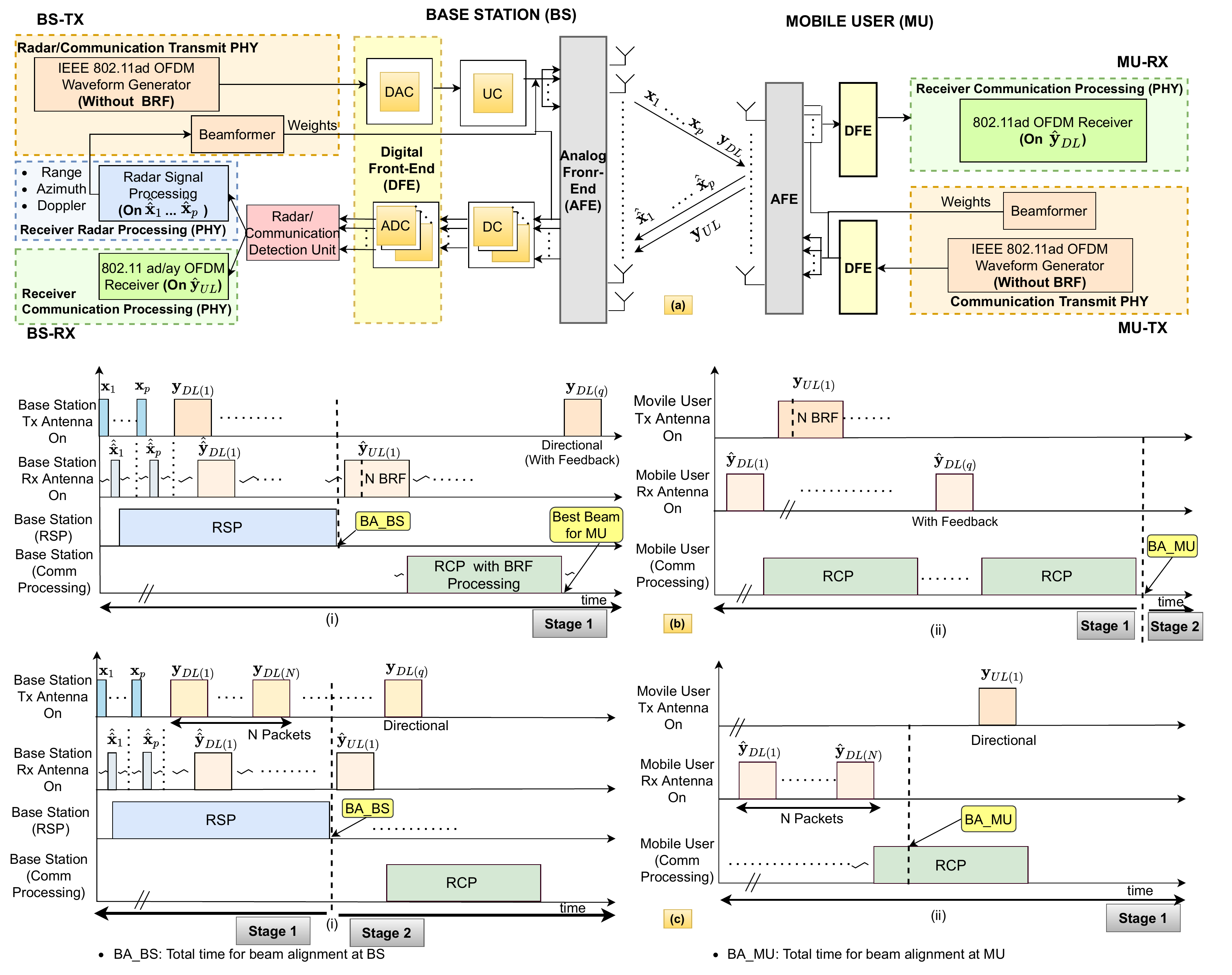}
    \vspace{-4.5mm}
    \caption{(a) Proposed JRC system architecture with beam alignment of BS through radar signal processing and corresponding (b) timing diagram of JRC \emph{Version-1} where beam alignment of MU is through longer uplink packet with multiple ($N$) BRF fields; (c) Timing for JRC \emph{Version-2} where beam alignment of MU is through reception of multiple ($N$) downlink packets through separate MU beams.} 
    \label{fig:JRC_arc}
    \vspace{-5.5mm}
\end{figure*}

Now discuss the proposed JRC framework at the BS, as shown in Fig.\ref{fig:JRC_arc}(a), to reduce the latency caused by the beam alignment procedure.
During the beam alignment \emph{Stage 1}, the transmitter at the BS-TX generates both radar ($\xp$) and communication waveforms ($\yDL$). The optional BRF training fields used in $\yDL$ for the standard are omitted here resulting in a shorter communication waveform.

In prior art in \cite{kumari_adaptive_2020,duggal2020doppler}, the radar waveform was embedded within the communication packet. However, there are several limitations to this signal model. First, short ranges typical of automotive environments ($\sim$40 to 200m), result in very short two-way propagation times (of the order of several nanoseconds), which fall within the duration of the $\yDL$ packet (which is of several microseconds). Hence, the signal model suggested in the prior art would require separate sets of mmW antennas and hardware to support simultaneous transmission of $\yDL$ through BS-TX while receiving the $\doublehat{\mathbf{x}}_p$ scattered from MU at BS-RX. Second, the system must manage the interference introduced by the strong transmission signal received directly at the receiver in order to be able to detect the weak radar scattered signal. Third, the reciprocity in the propagation channels for $\yDL$ and $\yUL$ communications would no longer hold, requiring separate beam alignment procedures for $\yDL$ and $\yUL$. 

For all these reasons, we propose that a separate radar waveform composed of just a section of the 802.11ad preamble is transmitted by BS-TX antennas through a quasi-omnidirectional beam. $P$ pulses of the radar waveform, denoted by $\xp, p = 1:P$,  are transmitted at a pulse repetition interval of $T_{PRI}$ as shown in the timing diagram presented in Fig.\ref{fig:JRC_arc}(b) and (c). Due to the short duty cycle of the radar pulse with respect to the $T_{PRI}$, the same mmW front-end can be used for both radar transmission at BS-TX and reception at BS-RX. 
The radar signals are upconverted from baseband to analog in the DFE and subsequently to mmW in the AFE. Since the radar waveform is composed of the 802.11ad packet preamble features, the same DFE and AFE and phased antenna array can be used for generating and transmitting both $\xp$ and $ \yDL$ without increasing the cost and complexity of the system hardware. The transmitted signal $\xp$ scatters at the MU and is received at the BS-RX where it is down-converted and digitized at the AFE and DFE. We introduce a radar-communication decision unit at the BS-RX for identifying scattered returns of $\doublehat{\mathbf{x}}_p$ from the scattered $\yDLss$ and uplink packet $\yULs$. The scattered $\doublehat{\mathbf{x}}_p$ are processed at the radar signal processing unit (RSP) where the MU are detected based on their range returns, identified from static clutter based on their Doppler and subsequently located in the angular space. This information from the RSP is used as feedback to the beamformer to enable the alignment at the BS-TX and BS-RX during \emph{Stage 2}. The total time required for beam alignment at BS is therefore the duration of the $P$ pulse repetition intervals ($P \times T_{PRI}$), and the radar signal processing time. Since these packets are considerably shorter than $\yDL$, the total duration for beam alignment is expected to be considerably lower than the standard. Note that in our proposed architecture, the RSP is introduced only in the BS and not in the MU. In other words, we have proposed digital beamforming at the BS-RX but retained analog beamforming at BS-TX, MU-TX, and MU-RX due to cost and complexity considerations. 
Hence, the beam alignment at the MU will be carried out as per the standard protocol. This architectural design choice is made since the number of beams at the BS are typically much greater than at the MU due to the higher number of antennas at the BS compared to the MU. We propose two versions for the beam alignment at the MU as shown in Fig.\ref{fig:JRC_arc}(b) and (c). \emph{Version-1} of the JRC architecture as shown in Fig.\ref{fig:JRC_arc}(b), is where the MU sends a $\yUL $ packet with $N$ BRF fields along $N$ pre-determined beams. Then the best beam is determined at the BS-RX and subsequently communicated to the MU through the next $\yDL$. Therefore, the total duration for beam alignment at the MU is based on the length of one long packet $\yUL$ with $N$ BRF, propagation delay, and processing times at BS and MU. 

\begin{algorithm*}
\small{
\caption{Beam alignment procedure for JRC}\label{alg:JRCv1_ta_steps}

\begin{steps}[leftmargin=1.5cm, label=Step \arabic*:]
  \item BS-TX transmits $P$ packets of $\mathbf{x}_p$ with $50\%$ duty cycle through quasi omnidirectional beams which are reflected from the MU and other targets in the environment.
  \item The received $P$ packets of $\doublehat{x}_p$ are identified at the radar/comm detection unit and  processed at BS-RSP to determine the azimuth of MU $(\theta_b)$. Meanwhile, BS-TX transmits quasi-omnidirectional $\mathbf{y}_{DL}$ packets without BRF fields.
  \item \textbf{Version-1:} The first $\hat{\mathbf{y}}_{DL}$ is received and processed at MU-RX (RCP). Then the MU-TX transmits $\mathbf{y}_{UL}$ with $N$ BRF fields.\\
   \textbf{Version-2:} The first $N$ received $\hat{\mathbf{y}}_{DL}$ packets are processed at MU-RCP and the correlation gains from preamble determines best beam, $\tilde{n}$. The duration of \emph{Stage-1} is based on the greater of the lengths of the two durations - RSP at BS and RCP at MU.
  \item\textbf{Version-1:} The received $\hat{\mathbf{y}}_{UL}$ is identified at the radar/comm detection unit and processed at BS-RCP when the best beam, $\tilde{n}^{th}$ is determined.
  \item \textbf{Version-1:} The next $\mathbf{y}_{DL}$ is transmitted without BRF fields along $\theta_b$ to MU along with information regarding $\tilde{n}^{th}$ BRF field.
  \item \textbf{Version-1:} The directional $\hat{\mathbf{y}}_{DL}$ is received and processed at MU-RCP to learn about $\tilde{n}^{th}$ BRF field. This completes \emph{Stage-1}.
  \end{steps}}
\end{algorithm*}
\emph{Version-2} of the JRC architecture as shown in Fig.\ref{fig:JRC_arc}(c) on the other hand, is similar to Fig.\ref{fig:std_arc}(b) where $N$ $\yDLs$ packets are received along $N$ beams at the MU to determine the best beam at MU. The duration of \emph{Stage 1} is based on the duration to determine the correlation gains of preambles of $N$ DL packets in \emph{Version-2} along with the propagation time and the communication processing time. Again, since the time complexity is of the order of $N$ rather than $M \times N$, we expect that the beam alignment will be quicker for the MU thereby supporting lower latency. The steps for beam alignment in \emph{Stage-1} for both versions are summarized in Algorithm\ref{alg:JRCv1_ta_steps}.

Thus the duration of the beam alignment stage for all three architectures is a function of the signal/packet lengths and the signal processing times. More details on the comparison of the timing values are provided in Section.\ref{sec:Results}A with suitable examples. 
\section{JRC Signal Model}
\label{sec:signalmodel}
In this section, we provide a detailed discussion of the JRC signal model and the corresponding software prototype required to generate the signal at the transmitter and process the signal at the receiver.
\subsection{JRC Transmitter Signal Model}
\begin{figure}[!b]
    \centering
    \includegraphics[scale=0.48]{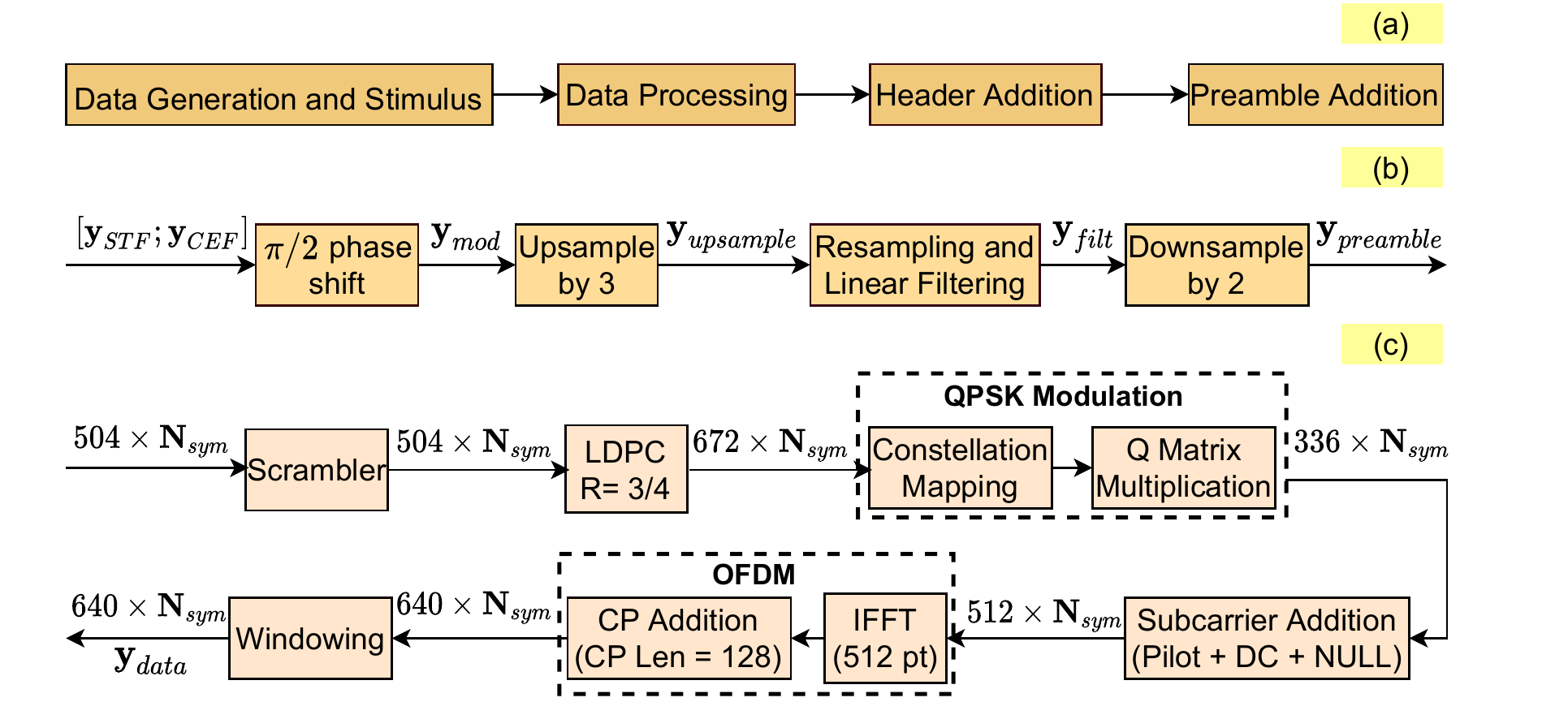}
    \vspace{-4mm}
    \caption{(a) IEEE 802.11ad JRC waveform generation; (b) preamble and BRF processing; and (c) header and data processing.}
    \label{fig:packetprocessing}
    \vspace{-2mm}
\end{figure}

The different blocks of the waveform generator within the transmitter are shown in Fig.\ref{fig:packetprocessing}(a). 
We will describe the generation of each part of the packet.

\textbf{Preamble and BRF field processing:} Both the preamble and BRF fields are always SC modulated even when header and data fields are modulated with OFDM. The preamble is divided into the short training field (STF) and the channel estimation field (CEF) as shown in Fig.\ref{fig:jrcwaveform}(b). The STF consists of sixteen repetitions of 128-bit Golay sequences denoted by $\mathbf{g}_{a_{128}}$ followed by one repetition of $-\mathbf{g}_{a_{128}}$. 
The STF bits are phase-shifted by integer multiples of $\pi/2$ as shown in -
\par\noindent\small
\begin{align}
    \mathbf{y}_{STF}[mT_c] =
    \begin{cases}
        \mathbf{g}_{a_{128}}[m\%128]exp(+j\frac{\pi}{2}m); \\
        \;\;\;\;\;\;\;\;\; m = 0,1,\cdots, 16\times 128 -1 \\
        -\mathbf{g}_{a_{128}}[m\%128]exp(+j\frac{\pi}{2}m); \\
        \;\;\;\;\;\;\;\;\; m = 16 \times 128 \cdots (17 \times 128)-1.
    \end{cases}
\end{align}\normalsize
Here, $T_c$ is bit period and it is reciprocal of the sampling frequency of 1.76GHz and $\%$ indicates a modulus operation. The CEF consists of a 512-bit Golay complementary pair, denoted by $\mathbf{g}_{u_{512}}$ and $\mathbf{g}_{v_{512}}$ followed by a 128 bit Golay sequence denoted by $\mathbf{g}_{v_{128}}$ which are similarly modulated to \par\noindent\small
\begin{align}
  \begin{split}  
    \mathbf{y}_{CEF}[mT_c] = &\left(\mathbf{g}_{u_{512}}[m] + \mathbf{g}_{v_{512}}[m-512] +\mathbf{g}_{v_{128}}[m-1024]\right)\\& exp(+j\frac{\pi}{2}m); m = 0, 1, \cdots 1151.
  \end{split}  
\end{align} \normalsize
Together, they are concatenated to form the total preamble $\mathbf{y}_{mod}$. 
The sampling frequency for OFDM data bits is 2.64GHz which is 1.5 times that of SC used for preamble data bits and hence up-conversion and down-conversion of the Golay sequences are required. First, the preamble bits are upsampled by a factor of three as shown below -
\par\noindent\small
\begin{align}
  \mathbf{y}_{upsample}\left[m\frac{T_s}{2}\right]= \begin{cases}
      \mathbf{y}_{mod}[m\frac{T_c}{3}]; & m = 0, 3, 6 \cdots\\
      0; & \text{otherwise}.
  \end{cases}  
\end{align} \normalsize
Here, $T_s$ is the sampling time corresponding to 2.64GHz. The up-sampled signal is subsequently smoothened by a linear digital filter where $\mathbf{h}_{filt}[k], k = 1\cdots K$ are the filter coefficients \cite{noauthor_ieee_2016-1} to obtain
\par\noindent\small
\begin{align}
\begin{split}
\mathbf{y}_{filt}\left[m\frac{T_s}{2}\right] = \sum_{k=1}^K \mathbf{y}_{upsample}\left[(m-k)\frac{T_s}{2}\right]\mathbf{h}_{filt}[k],\\
m = 0,1,\cdots 4992; k = 1, 2 \cdots K.
\end{split}
\end{align}\normalsize
Then the signal is downsampled by a factor of 2 to obtain
\par\noindent\small
\begin{align}
\begin{split}
    \mathbf{y}_{preamble}[mT_s]= \mathbf{y}_{filt}\left[2m\frac{T_s}{2}-\frac{K-1}{2}\frac{T_s}{2}\right]\\
   m = 0,1, \cdots 4992.
\end{split}
\end{align}\normalsize
The total length of $\mathbf{y}_{preamble}$ is 4992 and it is of 1.89$\mu s$ duration. The steps for generating the preamble bits are summarized in Fig.\ref{fig:packetprocessing}(b). The radar waveform is generated using the CEF bits as indicated in Fig.\ref{fig:jrcwaveform}(b). Hence each $p^{th}$ frame of the radar waveform corresponds to a uniquely generated $\xp=\mathbf{y}_{preamble}[4033:4800]$ consisting of $M' = 768$ samples. Since radar bits are already sampled at 2.64GHz, there is no need for sample rate conversion. Note that the radar waveform across consecutive $T_{PRI}$ are not identical as they generated from different seeds used for generating Golay complementary sequences. 

The BRF fields are optional fields to be used when beam alignment is required. Each BRF consists of two sub fields - automatic gain control (AGC) and training (TRN) fields - as shown in Fig.\ref{fig:jrcwaveform}(c). The AGC subfields consist of 64 bit Golay sequences, $\mathbf{g}_{a_{64}}$, repeated $M$ times where $M$ is the total number of beams supported by analog beamforming for the phased array in BS-TX. The BRF fields consist of the CEF followed by complementary 128 bit Golay sequences, $\mathbf{g}_{a_{128}}$ and $-\mathbf{g}_{b_{128}}$, arranged as shown in Fig.\ref{fig:jrcwaveform}(c) and replicated $M/4$ times. 
The CEF in the BRF is identical to the the CEF in the preamble. The bits in the BRF are generated at a chip duration of $T_c$ and subsequently processed in a manner identical to the preamble as shown in Fig.\ref{fig:packetprocessing}(b) to obtain $\mathbf{y}_{BRF}$. Their total duration can be calculated as ${2.845 \times [(M/4)]\mu} s$. In the proposed JRC architecture, these fields are not needed and skipped during \emph{Stage 1}. In the standard architecture, these are used in \emph{Stage 1} but skipped in \emph{Stage 2}.

\textbf{Header and data processing:} 
The 802.11ad standard specifies that data and header are modulated over 336 data subcarriers per OFDM symbol and transmitted over a minimum of 20 OFDM symbols if the optional BRF fields are used. In our proposed JRC, the number of OFDM symbols, $N_{sym}$, will be based on the amount of data to be transmitted as BRF are omitted. An additional OFDM symbol is used (for both standard and JRC) to specify the control information in the header such as amount of data transmitted, and information regarding the modulation and coding schemes (MCS) which must be subsequently used at the BS-RX and MU-RX to recover the data in $\yDLs$ and $\yULs$ respectively. Based on the specifications, the header must be modulated with quadrature phase shift keying with $3/4$ coding rate ($1$ parity bit for every $3$ data bits). The standard supports the BPSK, QPSK, 16 bit QAM and 64 bit QAM modulations, and $1/2$, $3/4$, $5/8$, and $13/16$ coding rates for the data. In this work, we use the same MCS for both data and header for simplicity. The header and data are processed as shown in Fig.\ref{fig:packetprocessing}(c). Two data bits are mapped to each of the 336 data subcarriers corresponding to a single OFDM symbol. At a $3/4$ coding rate, this corresponds to $3/4 \times 672 = 504$ bits per OFDM symbol per packet. If the number of data bits are less than an integral multiple of $504$, appropriate zero padding is done. Each block of $504$ data bits is sequentially passed through a scrambler which is a linear feedback shift register. The first seven bits of the header provide a unique scrambler initialization key to enable bit recovery at the receiver. The scrambled bits are then passed through a low density parity check (LDPC) encoder with the $3/4$ code rate. The standard specifies a parity check matrix, $\mathbf{\mathcal{H}}$, that operates upon the code word $\mathbf{c}$ formed of data bits followed by parity bits of a single OFDM symbol such that $\mathbf{\mathcal{H}}\mathbf{c^T}=0$. After LDPC encoding, each frame has 672 bits which are then modulated into the QPSK WLAN constellation followed by multiplication with a unit matrix, termed as $Q$ matrix \cite{noauthor_ieee_2016-1}, to randomise the amplitude of the modulated bits. The outcome of the operations are the 336 modulated complex samples per OFDM symbol, to which additional DC, NULL and pilot symbols are added, resulting in 512 complex symbols. Next, OFDM modulation is performed comprising of 512-IFFT and 128-length cyclic prefix (CP) addition. At the end,  weighted overlap-and-add (WOLA) windowing is performed on the OFDM modulated header and data to control the out-of-band emission \cite{agrawal_spectral_2020} to obtain $\mathbf{y}_{data}$. In the end, a scheduler combines the header and data with preamble, and BRF samples to form the packet which is then further processed by DFE and AFE.

Both the radar signal and the communication signals are converted from digital to analog as shown in
\par\noindent\small
\begin{align}
    \xp(t) = \sum_{m=0}^{M'}\xp[mT_s]\delta\left(t-mT_s\right),\\
    \yDL(t) = \sum_{m=0}^{M}\yDL[mT_s]\delta\left(t-mT_s\right).
\end{align}\normalsize
Here $M'=768$ (512 Golay samples upsampled by a factor of 1.5) indicates the number of samples within the radar signal which remains fixed since it is independent of data. The number of samples within $\mathbf{y}_{DL}$ varies based on the amount of data to be transmitted and the corresponding $N_{sym}$ OFDM symbols. These signals are then amplified such that energy $E_s$ is imparted to each symbol. The amplifier output is convolved with a root raised cosine transmit shaping filter, $\mathbf{h}_T$, and then upconverted to the mmW carrier frequency $f_c$ as shown in 
\par\noindent\small
\begin{align}
   \xp_{up}(t) = \sqrt{E_s}\left(\xp(t) \ast \mathbf{h}_T(t)\right)e^{+j 2\pi f_c t}\\
    \yDL_{up}(t) = \sqrt{Es}\left(\yDL(t)\ast \mathbf{h}_T(t)\right)e^{+j 2\pi f_c t}.
\end{align} \normalsize
Then the signals are transmitted through analog beamforming through an $N_{BS}$ element uniform linear array (ULA) after application of the antenna weight vector, $\mathbf{w}_{BSTX} \in \mathcal{C}^{N_{BS}\times 1}$, as shown in \par\noindent\small
\begin{align}
    \mathbf{X}_{p_{up}}(t) = \mathbf{w}_{BSTX}\xp_{up}^T(t),\\
    \mathbf{Y}_{{DL}_{up}}(t) = \mathbf{w}_{BSTX}\yDL_{up}^T(t).
\end{align}\normalsize
The above process is repeated till all the $P$ radar packets are transmitted followed by the communication packets after a guard time interval. In \emph{Stage-1}, the antenna weight vector (AWV) is chosen to support a quasi-omnidirectional beam. At the end of \emph{Stage-1}, when $\theta$ of the MU is determined, then the AWV is selected to support a directional beam along $\theta$. 
\subsection{Radar Signal Detection and Processing}
\label{Sec:RSP}
In this section, we first discuss the received radar signal model and then the radar signal processing in detail. The radar transmits signal $\mathbf{X}_{p_{up}}(t)$ is scattered by the dynamic MU and received at the $N_{BS}$ element ULA at the BS-RX. If we assume that the MU is a dynamic target with $B$ point scatterers, each at range $r_b(t)$, then the received signal at the BS-RX antenna array is
\par\noindent\small
\begin{align}
\label{eq:rx_radar}
    \doublehat{\mathbf{X}}_{p_{up}}(t) = \sum_{b=1}^B a_b \mathbf{u}_{\theta_b(t)}\mathbf{u}^T_{\theta_b(t)}\left[\mathbf{X}_{p_{up}}(t)\ast \mathbf{H}^2\left(\frac{2r_b(t)}{c}\right)\right] + \mathbf{Z}(t).
\end{align}\normalsize
Here $a_b$ captures the amplitude of the reflection at each point scatterer and $\mathbf{H}^2 = [\mathbf{h}^2(\frac{2r_1(t)}{c}) \cdots \mathbf{h}^2(\frac{2r_B(t)}{c})]$ is the two-way propagation factor which is a function of the time-varying range of each $b^{th}$ point scatterer. The steering vector from the antenna array to and from $\theta_b$ is given by $\mathbf{u}_{\theta_b}^T = [1\;\;e^{jkd_{BS}\sin \theta_b}\;\;\cdots]$ where $k$ is the propagation constant at $f_c$ and $d_{BS}$ is the uniform element spacing. We assume that the MU is within the maximum unambiguous range of the target i.e. $\frac{2r_b(t)}{c} < T_{PRI}$. $\mathbf{Z}$ is the additive circular symmetric white Gaussian noise at the $N_{BS}$ receiver chains. 
We assume that each point scatterer is moving with a constant radial velocity $v_b$, then $r_b(t) = r_{ob}-v_b(t)t$ where $r_{ob}$ is the initial distance from the BS. After down-conversion and digitization and in the absence of multipath, \eqref{eq:rx_radar} is
\par
\begin{figure*}[htbp]
    \centering
    \includegraphics[scale=0.95]{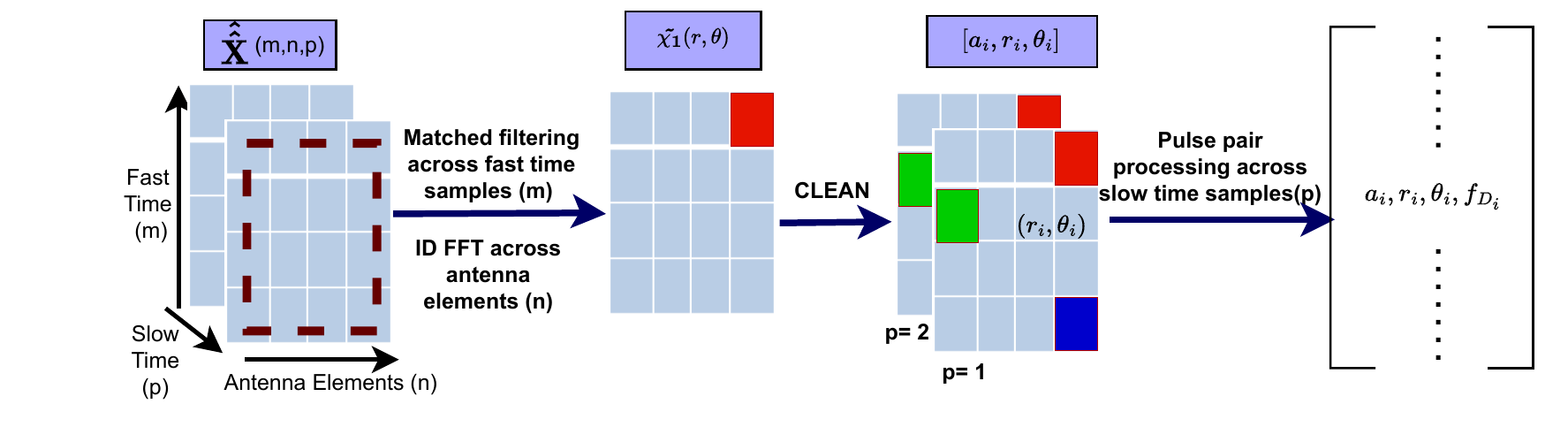}
    \vspace{-4mm}
    \caption{Radar signal processing block: Input radar data cube is first processed to obtain range-azimuth ambiguity plots. Multiple targets are detected using CLEAN. Corresponding Dopplers of each target are estimated through pulse-pair processing.}
    \label{fig:rsp_bd}
    \vspace{-4mm}
\end{figure*}

\noindent\small
\begin{align}
\label{eq:rx_radar2}
\doublehat{\mathbf{X}}_p[m] = \sum_{b=1}^B a_b \mathbf{u}_{\theta_b[m]}\mathbf{u}^T_{\theta_b[m]} \mathbf{X}_p\left[m-m_b\right]e^{-j2\pi f_{D_b}pT_{PRI}} + \mathbf{Z}[m],
\end{align}\normalsize
where $m_b$ is the sample index corresponding to the time delay ($\frac{2r_b(t)}{c}$) and $f_{D_b} = \frac{2v_bf_c}{c}$ is the Doppler frequency of the $b^{th}$ point scatterer. Note that the received signal in \eqref{eq:rx_radar2} forms a data cube across fast time ($m=1:M'$), slow time ($p=1:P$) and the $n=1:N_{BS}$ receiver elements.   
The radar received signal, $\doublehat{\mathbf{X}}_p[m]$ in \eqref{eq:rx_radar2}, is distinguished from uplink signal $\mathbf{Y}_{UP}[m]$ by the \textit{radar-communication detection unit} based on positive correlation with $\mathbf{X}_p[m]$ sequence. Note that Golay sequences in the $\doublehat{\mathbf{X}}_p$ will be distinct from the sequence in $\mathbf{Y}_{UP}$. Then, the data cube is processed within the RSP as shown in Fig.\ref{fig:rsp_bd} to estimate the range ($r_b$), azimuth ($\theta_b$) and Doppler velocity ($v_b$) of the MU. 
In BS-TX, the radar signal consisting of the Golay sequence was up-sampled from the original 1.76GHz to 2.64GHz to match the sampling rate of the OFDM modulated data. Hence, the first step in the RSP of the BS-RX is to down-sample the fast time data from each $p^{th}$ packet and $n^{th}$ antenna, by a factor of 1.5, to 1.76GHz (from 768 samples to 512 samples) to obtain $\doublehat{\mathbf{X}}_{p_{ds}}[m]$. The fast time samples from each RF chain of the first packet (shown with the red dotted rectangle in Fig.\ref{fig:rsp_bd}) are cross-correlated with the corresponding Golay sequence as shown in
\par\noindent\small
\begin{align}
   \mathbf{\chi}_1[r] = \doublehat{\mathbf{X}}_{1_{ds}}[m] \ast \mathbf{G}_{u_{512}}[m], m = 1:512,
\end{align}
\normalsize
where $\mathbf{G}_{u_{512}} \in \mathcal{R}^{512 \times N_{BS}}$ consists of the $N_{BS}$ copies of the column vector $\mathbf{g}_{u_{512}}$. The cross-correlation operation provides a peak at $m_b$ corresponding to the range of the $b^{th}$ point scatterer provided the SNR is sufficient. The range axis spans 512 bins with a range resolution of $cT_c/2$. 
Then one-dimensional Fourier transform is carried out after zero-padding across the $N_{BS}$ elements to obtain the range-azimuth ambiguity plot $\tilde{\chi}_1[m,\theta] = \text{DFT} \{\chi_1\}$. The azimuth axis spans from $-90^{\circ}$ to $90^{\circ}$ (if the antenna spacing is half-wavelength) with a bin size governed by the zero-padding factor. The MU (one or multiple) are identified based on peaks in the range-azimuth ambiguity plot.  
In order to detect multiple targets across a large dynamic range, the CLEAN algorithm is implemented \cite{ram2007human,ram2008through}. This is an iterative algorithm, wherein in each $i^{th}$ step the strongest target is identified based on the peak, $a_i$, in the 2D ambiguity plot \small $\underset{r_i,\theta_i}{\arg\max}||\tilde{\chi}^i[r,\theta]||$. \normalsize
Then the 2D point spread response, $\Upsilon[r_i,\theta_i]$, of the peak target at this position is removed from the ambiguity plot to generate a residual plot \small
$\tilde{\chi}^{i+1}_1 = \tilde{\chi}^i_1-\Upsilon[r_i,\theta_i]$,
\normalsize
that is utilized for the subsequent step.
The above two steps are iterated till the strength of the residue plot falls below a predefined threshold. The output of the CLEAN algorithm is the list of possible targets, $i=1:I$, determined in each $i^{th}$ step of the CLEAN algorithm and their corresponding amplitudes and 2D positions ($a_i, r_i,\theta_i$). Target returns that fall very close to each other in the range-azimuth ambiguity plot possibly arise from a single extended target and are clustered together. Note that some of the targets may  not be MU and may be static clutter in the BS environment. Hence, we next implement Doppler processing on the returns. While many algorithms may be used for estimating Doppler velocity \cite{lee2018weighted,kumari_ieee_2018}, we implement the computationally efficient pulse pair processing which requires only two consecutive radar packets and is performed only for those cells in the range-azimuth ambiguity plots corresponding to the localized targets as shown in 
\begin{align}
f_{D_i} =  \frac{-1}{2\pi T_{PRI}} \arg\left[ \tilde{\chi}_2[r_i,\theta_i]\ast\tilde{\chi}_1[r_i,\theta_i]\right].
\label{eq:pulse_pair1}
\end{align}
Based on estimated Doppler $f_{D_b}$, we determine if the $i^{th}$ target is static or dynamic. If it is dynamic, then it is regarded as an MU and its $\theta_i$ is used to determine the antenna weight vector $\mathbf{w}_{BSTX}$ in the subsequent \emph{Stage-2} as discussed in Section~\ref{sec:SysArchitecture}.
\subsection{Communication Received Signal Processing }
The aim of the communication receiver is to demodulate and decode the transmitted data correctly. It performs the same tasks as the communication transmitter but in reverse order. The transmitter output propagates from BS-TX to the MU-RX at range $r$ and azimuth $\theta$ through the steering vector $\mathbf{u}_{\theta}^T$. If the MU-RX consists of $N_{MU}$ element ULA, then the received signal at the MU-RX after analog beamforming with antenna weight vector, $\mathbf{w}_{MURX}\in \mathcal{C}^{N_{MU}\times 1}$, is \par\noindent\small
\begin{align}
\hat{\mathbf{y}}_{{DL}_{up}}(t) = \mathbf{w}_{MURX}^T \mathbf{u}_{\phi} \mathbf{u}_{\theta}^T\left[\mathbf{Y}_{{DL}_{up}}(t)\ast \mathbf{h}^1\left(\frac{r(t)}{c}\right)\right] + \mathbf{\zeta}, 
\label{eq:rx_comm}
\end{align} \normalsize
where $\mathbf{u}_{\phi}^T = [1\;e^{jkd_{MU}\sin \phi}\;\cdots ]$ is the steering vector for the BS at $\phi$ with respect to MU-RX for $d_{MU}$ antenna element spacing, $\mathbf{h}^1(\frac{r(t)}{c})$ is the one-way propagation factor for each MU and $\zeta$ is the additive circular-symmetric white Gaussian noise.
\begin{figure}[!t]
    \centering
    \includegraphics[scale=0.52]{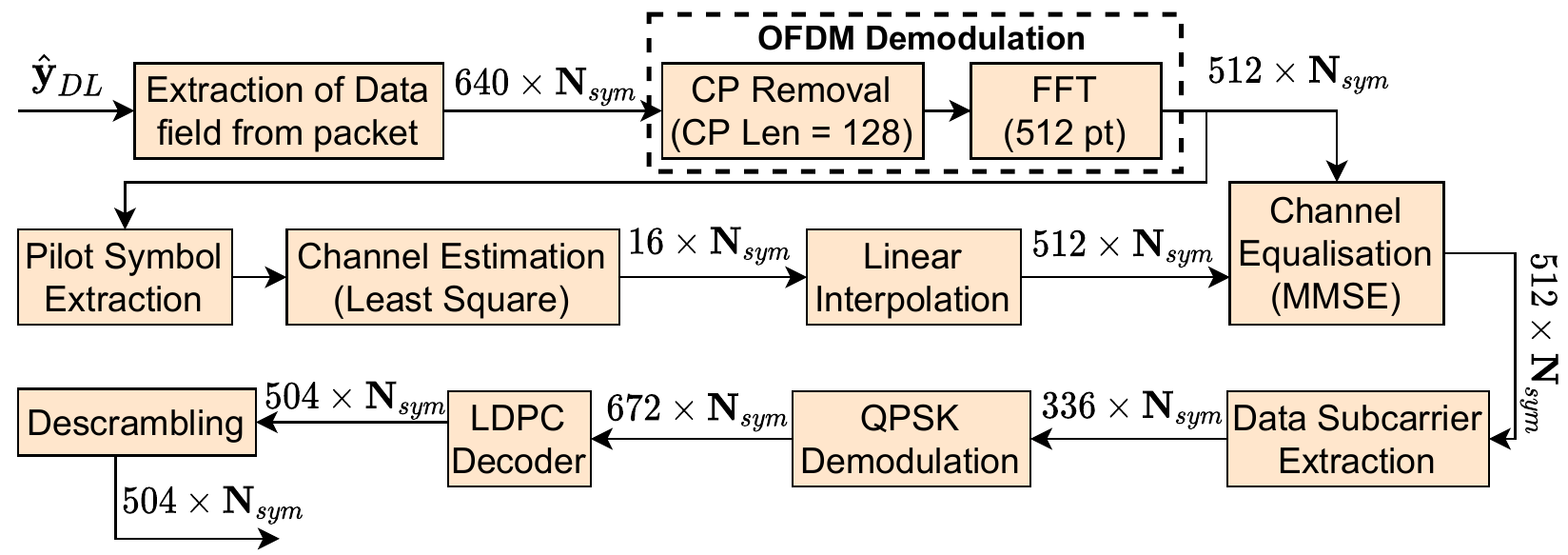}
    \vspace{-2mm}
    \caption{Data recovery at communication receiver}
    \label{fig:rx_bd}
    \vspace{-3mm}
\end{figure}
The received RF signal is amplified, then down-converted and digitized before it is further processed through the blocks shown in Fig.\ref{fig:rx_bd} to extract data from $\hat{\mathbf{y}}_{DL}$. 

The first step is to perform OFDM demodulation which consists of cyclic-prefix removal followed by 512-FFT. Then, least-squares-based channel estimation is performed using the extracted pilots. 
Then, minimum mean square error \cite{ibnkahla2004signal} based channel equalization is carried out on the 336 data symbols in each OFDM symbol. This is followed by soft data demodulation \cite{viterbi1998intuitive,olivatto2017simplified} and LDPC channel decoding \cite{liu2014variable}.The output of the LDPC decoder is passed through the descrambling block, where a dummy header of 57 bits is added and padded bits are removed and passed through the same scrambling block as the one used in the transmitter. The output after descrambling is the decoded data bits extracted from $\hat{\mathbf{y}}_{DL}$. Similar processing is also done at BS-RX to recover the data transmitted over the uplink $\hat{\mathbf{y}}_{UL}$ .
\section{Simulation Setup}
\label{sec:simsetup}
In this section, we describe the simulation setup for evaluating the communication link metrics and the radar performance metrics. 
\subsection{Target Model}
\begin{figure}[!ht]
    \centering
    \includegraphics[scale=0.6]{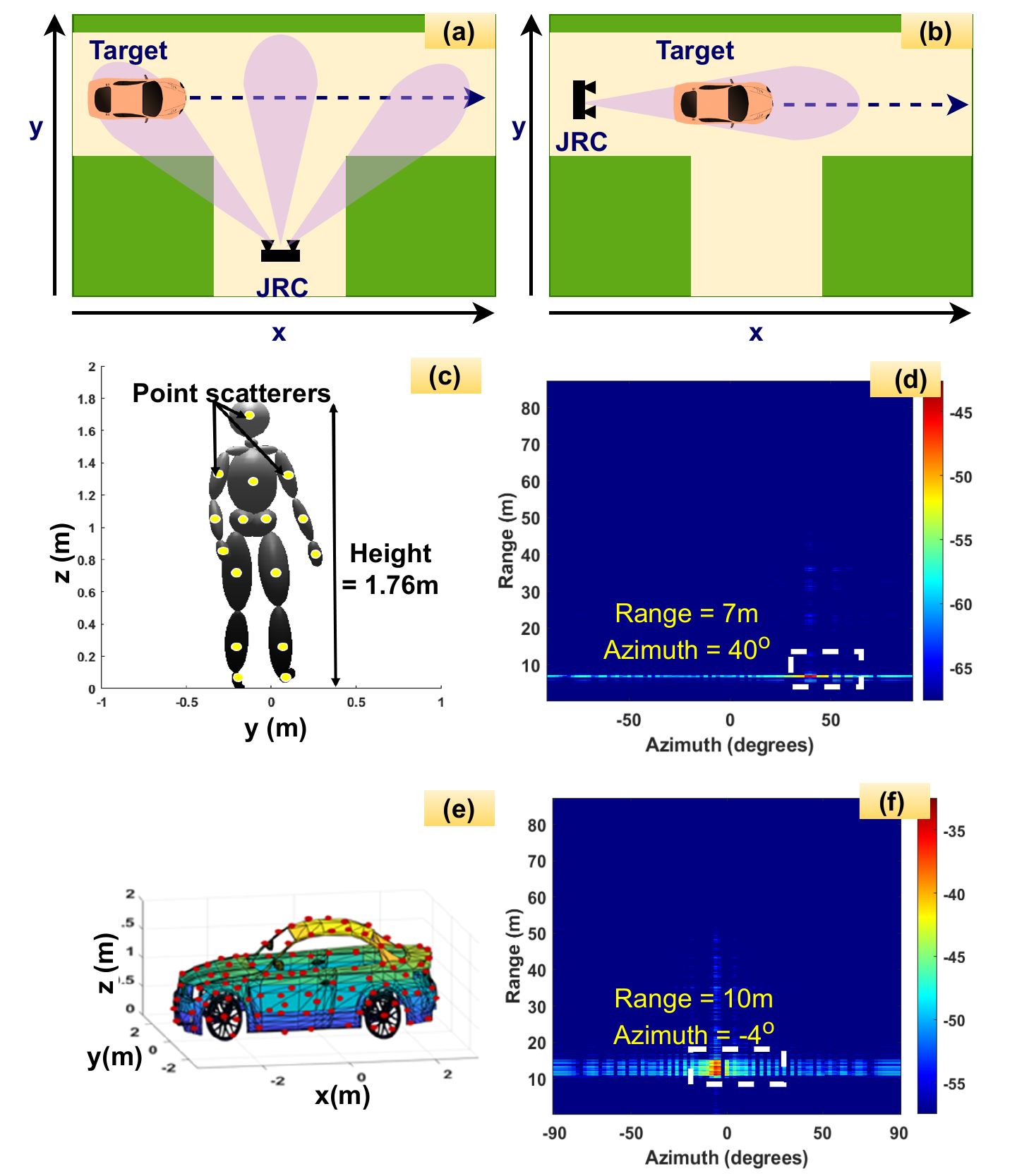}
    \vspace{-3mm}
    \caption{Trajectories followed by a target/MU with respect to JRC: (a) tangential motion (b) Radial motion. (c) Point scatterer model of pedestrian (MU), and (d) 2D range-azimuth ambiguity plot for pedestrian. (e) Point scatterer model of a mid-size car (MU) and (f) 2D range-azimuth plot for the car.}
    \label{fig:sim_set}
    \vspace{-5mm}
\end{figure}

We consider a three-dimensional Cartesian space where $xy$ forms the ground and $z$ is the height axis. The target motion along two trajectories over a duration of 1s is shown in Fig.\ref{fig:sim_set}(a) and (b). For the first trajectory, we assume that the JRC is located at $[0m,0m,0m]$ and the target starts from $[-5m,10m,0m]$ and moves along the $x$-axis in a direction tangential to the JRC. For the second trajectory, the JRC is assumed to be at $[-13m,15m,0m]$ while the target starts at $[-3m,15m,0m]$ and moves along the $x$ axis in a radial direction away from the JRC. We consider two types of dynamic target models/MU - the first is a simple isotropic point target and the second is an extended target with multiple point scatterers and aspect angle dependence. We consider two extended targets - a pedestrian and a mid-size car. For both trajectories, the pedestrian moves with a constant velocity of 1.5m/s while the car moves with a velocity of 10m/s. The animation data for the walking pedestrian are obtained from motion capture technology \cite{ram2008simulation,ram2010simulation}. The data describes the dynamics of a collection of 27 body parts of a skeleton at a frame rate of 60Hz. Each body part is modeled as an ellipsoid, as shown in Fig.\ref{fig:sim_set}(c), whose radar cross-section is obtained through analytical expressions.
Based on the methods described in the previous section, we generate the range-azimuth ambiguity plot of the human and display the result for one time frame in Fig.\ref{fig:sim_set}(d). The figure shows that the human is essentially a spatially small target and the point scatterers are clustered together within a small region of the ambiguity plot. 
Our second extended target is a mid-size car modelled using a 3D computer-aided design. The car is animated along the above mentioned trajectories based on the methodology described in \cite{pandey2020database,pandey2022classification}. The metallic parts of the mid-size car are rendered into $6905$  triangular facets and their 3D centroids are identified as the collection of point scatterers as shown in Fig.\ref{fig:sim_set}(e). The corresponding range-azimuth ambiguity plot for one timeframe is displayed in Fig.\ref{fig:sim_set}(f). The result shows that the car is a larger spatial target than the pedestrian and thus results in a slightly wider spread along range and azimuth. The centroid of the car returns is determined by clustering. In both cases, the Dopplers were non-zero, indicating that these are dynamic targets and not static clutter. 
\vspace{-2 mm}
\subsection{Channel Model}
We consider two types of channel models to model the two-way radar propagation ($\mathbf{h}^2$) and one-way communication propagation ($\mathbf{h}^1$) channels. First, we consider an idealistic free space channel model with a direct line-of-sight (LOS) between the BS and MU and no multipath/non-line-of-sight (NLOS). The one-way propagation path between the BS and MU which are $r$ distance apart is given by \par\noindent\small
\begin{align}
  \mathbf{h}^1\left(\frac{r(t)}{c}\right) = \sqrt{\frac{G_{BS}(\theta)G_{MU}(\phi)L_f(r)\lambda^2}{(4\pi r(t))^2}}\delta\left(t-\frac{ r(t)}{c}\right),
  \label{eq:one_way_free}
\end{align}\normalsize
where $G_{BS}$ and $G_{MU}$ are the element patterns of the BS and MU antenna arrays, $L_f(r)$ is the atmospheric attenuation and $\delta(\cdot)$ is the Dirac-delta function. Similarly, the two-way propagation from the BS to the MU and back, for the radar signal is shown as \par\noindent\small
\begin{align}
  \mathbf{h}^2\left(\frac{2r(t)}{c}\right) = \sqrt{\frac{G_{BS}^2(\theta(t))L_f(2r)\lambda^2}{(4\pi)^3 r^4(t)}}\delta\left(t-\frac{2 r(t)}{c}\right).
\end{align}\normalsize

Second, we consider the \emph{Rician} fading model which has been popularly used to model outdoor mmW communication \cite{yan2020channel}. Here, the propagation is characterised by a dominant LOS component and additional NLOS components. The communication signal received at MU-RX shown in \eqref{eq:rx_comm} is modified for the fading channel model as
\par\noindent\small
\begin{align}
\begin{split}
  \hat{\mathbf{y}}_{{DL}_{up}}(t) =  \mathbf{w}_{MURX}^T \mathbf{u}_{\phi} \mathbf{u}_{\theta}^T\left[\mathbf{Y}_{{DL}_{up}}(t)\ast \sqrt{\frac{\ricJ}{\ricJ+1}}\mathbf{h}^1\left(\frac{r(t)}{c}\right)\right] \\
  +\left[\mathbf{\rho}_{nlos}(t)\ast \sqrt{\frac{1}{\ricJ+1}}\mathbf{h}^1\left(\frac{r(t)}{c}\right)\right]+\mathbf{\zeta},
\end{split}
\label{eq:Rician}
\end{align}
\normalsize
where $\ricJ$ is a Rician factor that is empirically determined for different propagation environments such as rural, urban and suburban \cite{zhu2014probability}. The first term in \eqref{eq:Rician} models the direct path in the one-way propagation factor while $\mathbf{\rho}_{nlos}(t)$ denotes the signal arising from the multipath that is modeled by an independent and identically distributed complex Gaussian random variable with zero mean and unit variance. The fading channel for two-way propagation of the radar signal $\doublehat{\mathbf{X}}_{p_{up}(t)}$ is modeled by squaring the one-way propagation model in \eqref{eq:Rician} for each antenna element at the BS-RX. 
\section{Numerical Results}
\label{sec:Results}
\subsection{Timing Analysis }
In this section, we compare the time taken for beam alignment - the duration of \emph{Stage-1} - for the standard IEEE 802.11ad and the two versions of the proposed JRC. The timing analysis is carried out based on the architecture presented in Section.\ref{sec:SysArchitecture} and the results are reported in Table.\ref{tab:timing_analysis}.
\begin{table*}[htbp]
\addtolength{\tabcolsep}{-3pt}
\centering
\caption{Timing analysis for beam alignment for Standard, JRC \emph{Version-1}, and JRC \emph{Version-2} }
\begin{tabular}{p{0.15\textwidth}|p{0.2\textwidth}|p{01.1cm}p{0.9cm}p{0.8cm}|p{1.1cm}p{0.9cm}p{0.8cm}|p{1.1cm}p{0.9cm}p{0.8cm}}
Signals  & Signal Processing                                            & \multicolumn{3}{c|}{Standard}      & \multicolumn{3}{c|}{JRC \emph{Version- 1}} & \multicolumn{3}{c}{JRC \emph{Version-2}} \\ \hline \hline
&   & Duration & Packets & Total & Duration    & Packets & Total  & Duration  & Packets  & Total \\ \hline
DL Packet $\mathbf{y}_{DL}$ (in $\mathbf{\mu}s$)         &                                                              & 29.73   & 4    & 118.9 & 4.5  & 1  & 4.5  & 4.5 & 4 & 18  \\
UL Packet $\mathbf{y}_{UL}$ (in $\mathbf{\mu}s$)         &     & 4.5 & 1  & 4.5   & 9.82 & 1   & 9.82 & -                & - & -             \\
Comm Idle Time (in $\mathbf{\mu}s$)   &   & 1 & 3 & 3     & -   & -  & - & 1  & 3   & 3  \\
Radar Waveform $\mathbf{x}_{Dp}$ (in $\mathbf{\mu}s$)    &    & -        & -              & -     & 0.29     & 2 & 0.58  & 0.29  & 2  & 0.58          \\
Pulse Repetition Interval $T_{PRI}$ (in $\mathbf{\mu}s$) &     & -        & - & -     & 0.29  & 2   & 0.58   & 0.29   & 2    & 0.58  \\ & RCP on $\hat{\mathbf{y}}_{DL}$ (in ms) & 16       & 4  & 64    & 16   & 2   & 32   & & -    & - \\
& RCP on $\hat{\mathbf{y}}_{UL}$ (in ms)  & 16   & 1              & 16    & 16   & 1    & 16 & - & - & -\\
& Receive Preamble Correlation $\hat{\mathbf{y}}_{DL}$ (in ms) & -        & -              & -     & -                & -   & -   & -   & -  & 4.5  \\
& Radar/Comm Detection Unit (in ms)    & -        & -              & -     & 2.5              & 2     & 5   & 2.5              & 1   & 2.5   \\  
& Preamble Bits Extraction (in ms)    & -   & -  & -     & 1.6 & -  & 1.6  & 1.6  & -   & 1.6   \\
& RSP on $\doublehat{\mathbf{x}}_{p}, p =1,2$ (in ms) & -   & - & - & 16 & - & 16  & 16 & -  & 16  \\ \hline
\multicolumn{2}{p{0.35\textwidth}|}{Total time for beam alignment at BS }& \multicolumn{3}{r|}{64+16+0.10=80.1}    & \multicolumn{3}{r|}{16+2.5+1.6 =20.1} & \multicolumn{3}{r}{16+2.5+1.6 =20.1}\\
\multicolumn{2}{p{0.35\textwidth}|}{Total time for beam alignment at MU }  & \multicolumn{3}{r|}{64 ms}& \multicolumn{3}{r|}{16+16+2.5=34.5}  & \multicolumn{3}{r}{4.5 ms}   \\
\multicolumn{2}{p{0.35\textwidth}|}{Total time for completion of  \emph{Stage 1}}& \multicolumn{3}{r|}{80.1 ms} & \multicolumn{3}{r|}{34.5 ms} &\multicolumn{3}{r}{20.1 ms}      
\\ \hline 
\end{tabular}
\label{tab:timing_analysis}
\end{table*}
For both versions of the JRC and the standard architectures, we assume that the BS and MU have a ULA of 32 antennas and 4 antennas respectively. The total beam alignment time is based on the number of packets, their duration as well as the processing time of packets which are of the order of milliseconds. 
In this work, all the processing is carried out in floating-point in MATLAB 2021a on an Intel Core i7 processor with 128 GB RAM. The one-way and two-way propagation times for communication and radar are neglected since they are comparably much shorter (of the order of nanoseconds) over the short ranges of the MU with respect to the BS than the signal lengths and signal processing times. 

In the standard protocol, $\mathbf{y}_{DL}$ consists of 32 BRF fields in addition to the preamble, header and data to support beam alignment at the BS. This results in a packet length of 29.73$\mu s$ duration for $N_{sym} = 20$ data symbols which is the minimum specified by the protocol. This packet is transmitted $4$ times with an inter-packet guard spacing of 1$\mu s$ to support beam alignment at the MU.  This information is then communicated through the subsequent $\mathbf{y}_{UL}$ transmitted over the MU's best beam to the BS where it is processed. Then, the next $\mathbf{y}_{DL}$ is transmitted over the BS's best beam and received by the MU's best beam in \emph{Stage-2}. The overall beam alignment duration for BS and MU are 80.1ms and 64ms respectively. The \emph{Stage-1} alignment time is computed by the total time required for both BS and MU beam alignment (the greatest value of the two times). Note that in the standard beam alignment, the overall duration is a function of the communication packet length and hence the number of OFDM data symbols. Also, the packets are long since they must include the BRF fields for all possible beams from the BS and repeated for all possible beams of the MU. The recovery of data bits from the received data packet is computationally intensive and hence requires a long time. After the data bit recovery from each packet, the BRF fields are processed for the correlation gains, generating an $M\times N$ matrix. Finally, the $(\tilde{m},\tilde{n})^{th}$ element of the matrix with the highest correlation gain where $\tilde{m}$ and $\tilde{n}$ provide the best beam among the $M$ BS beams and $N$ MU beams respectively. All the above factors result in a long beam alignment phase.
\begin{figure*}[!b]
    \centering
    \includegraphics[scale=0.475]{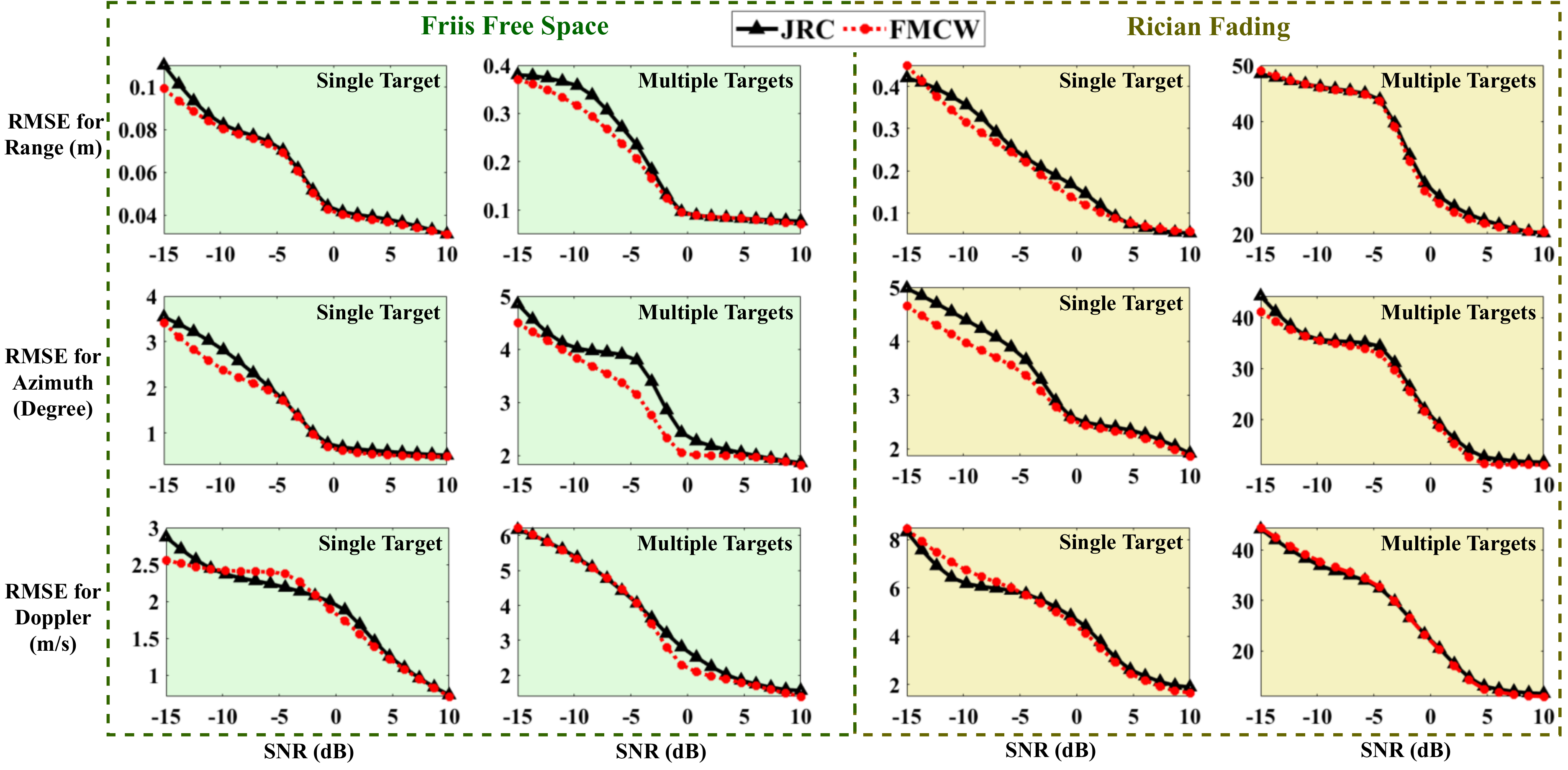}
    \vspace{-5mm}
    \caption{RMSE results for range, azimuth and Doppler velocity estimates from Monte-Carlo based numerical experiments of single and multiple point targets/MU in free space and Rician propagation conditions with JRC and FMCW radar.}
    \label{fig:rmse_point}
    \vspace{-4mm}
\end{figure*}
Next, we consider the \emph{Stage-1} beam alignment duration of the JRC. The two symbols of radar waveform ($\mathbf{x}_1$ and $\mathbf{x}_2$) are each of very short duration (0.29$\mu s$) since they consist of only 512 samples and are transmitted ahead of a $\mathbf{y}_{DL}$ packet with $N_{sym}=10$ symbols of data (without BRF fields). The $T_{PRI}$ for each radar signal is 0.58$\mu s$ which results in a maximum detectable range of approximately 87m. The reflected received signals, $\doublehat{\mathbf{x}}_p, p = 1,2$, are detected in the radar-communication detection unit and subsequently processed to estimate the MU’s range, Doppler, and azimuth angle. The azimuth angle of arrival forms the beam direction for the subsequent transmission of packets in \emph{Stage-2}. Note that the beam is directed exactly at the most recent position of the target, unlike the standard where the transmission in \emph{Stage-2} is along the optimum pre-determined beams. Hence the total time required at the BS for beam alignment is 20ms. This time requirement is entirely independent of the number of OFDM data symbols that are transmitted in $\mathbf{y}_{DL}$. Further, since the BRF fields are omitted in $\mathbf{y}_{DL}$, the minimum number of OFDM data symbols that must be transmitted in the packet is no longer 20 and instead directly based on the amount of data that is to be transmitted. The JRC is only included in the BS and not in the MU. Hence, we present two different versions of beam alignment at the MU.

In JRC \emph{Version-1}, $\mathbf{y}_{DL}$ is first received at MU with a quasi-omnidirectional beam. Once the data are processed, the MU transmits the $\mathbf{y}_{UL}$ with four BRF fields, such that each field is transmitted through each of the 4 different pre-determined beams of the MU. The received $\hat{\mathbf{y}}_{UL}$ is processed at the BS to determine the best $\tilde{n}^{th}$ beam of the MU. This information is communicated to the MU through the next $\mathbf{y}_{DL}$. Hence, the total duration for beam alignment at MU is 34.5ms which is shorter than the standard. In JRC \emph{Version-2}, the BS sends four consecutive $\mathbf{y}_{DL}$ packets. Each of these packets consists of 10 OFDM symbols and zero BRF fields. The MU receives each packet through the corresponding 4 pre-determined beams, one at a time. All four data packets are processed to estimate the best $\tilde{n}^{th}$ beam at the MU and the time taken for beam alignment at MU is 4.5ms. The short processing time is because the best beam is determined using the parallel correlation operation on the Golay sequences of the preamble of the multiple packets. From this timing analysis, it is evident that the JRC-based architecture provides faster beam alignment than the standard for the same number of beams at BS and MU. This can be attributed to two factors - first, the reduction of the packet duration due to the omission of the BRF fields due to radar-based beam alignment at the BS. This also reduces the total number of packets that have to be transmitted to support the beam alignment at the MU (especially in \emph{Version-1}). Second, digital beamforming and subsequent radar signal processing at the BS-RX support the detection and localization of the MU. This becomes even more advantageous when there are multiple MU in the channel since the above beam refinement protocol for multiple MU can be carried out in parallel since the radar can detect multiple targets. This is not possible with the standard protocol. 
\subsection{Radar Metrics}
In this section, we evaluate the performance of the 802.11ad-based radar and then benchmark it with the frequency modulated continuous wave (FMCW) which is commonly used in mmW radars. The FMCW radar signal parameters such as carrier frequency ($f_c$), sampling time ($T_{s}$), PRI ($ T_{PRI}$), and the number of radar pulses are selected to be identical to that of the JRC waveform. The chirp factor for the FMCW waveform is 600MHz/$\mu s$.  

First, we perform Monte-Carlo simulations with 10000 iterations for different radar target scenarios at various SNRs. We consider single and multiple target cases under free space and Rician propagation conditions. The value of ($\ricJ$) is chosen as 7dB in our work to model the channel according to the rural environment \cite{zhu2014probability}. For the single target scenario, in each iteration of the Monte-Carlo simulation, we consider the point target's location to be uniformly distributed in the 2D Cartesian space within 60m $\times$ 60m while the velocity of the target is varied uniformly from -30m/s to 30m/s. The RCS is varied based on the Swerling-1 model with the mean RCS selected randomly from 10m$^2$, 1m$^2$, 0.1m$^2$ and 0.01m$^2$. In the multi-target scenario, the number of targets for each iteration is varied based on the Poisson distribution with the average number of targets taken as 2. A target is localized through radar signal processing to provide range, azimuth, and Doppler estimates. We calculate the accuracy of these estimates by calculating the root mean square error (RMSE) of the estimates of the target with the ground truth as shown in Fig. \ref{fig:rmse_point} for different SNR. 

In both free space and Rician conditions, we observe that the  RMSE of the estimates decreases with an increase in the SNR for both JRC and FMCW waveforms. For the single target conditions, the lowest RMSE at the highest SNR corresponds to the resolution sizes of the estimates. In other words, the RMSE for range estimation is the range resolution which is a function of the signal bandwidth, while the RMSE for azimuth estimation is governed by the aperture size at the BS. Even though the Doppler velocity is estimated through a very short coherent processing interval of just two radar packets, the Doppler velocity estimate shows a low error because the targets were first resolved in range and azimuth before the Doppler processing was carried out. For all three parameters - range, azimuth, and Doppler - the error estimates for the JRC waveform are comparable with that of the FMCW waveform. Next, we consider the multiple targets scenario for the free space conditions. Here, we observe that the errors have  increased especially when the SNR is poor. The errors for both single and multiple targets under Rician conditions are substantially  worse than the free space conditions. This is especially true for the multiple target scenario under poor SNR. However, it is important to note that the performance of the JRC waveform is not any worse than that of the conventional FMCW under these circumstances. 

\begin{figure}
\centering
    \includegraphics[scale=0.475]{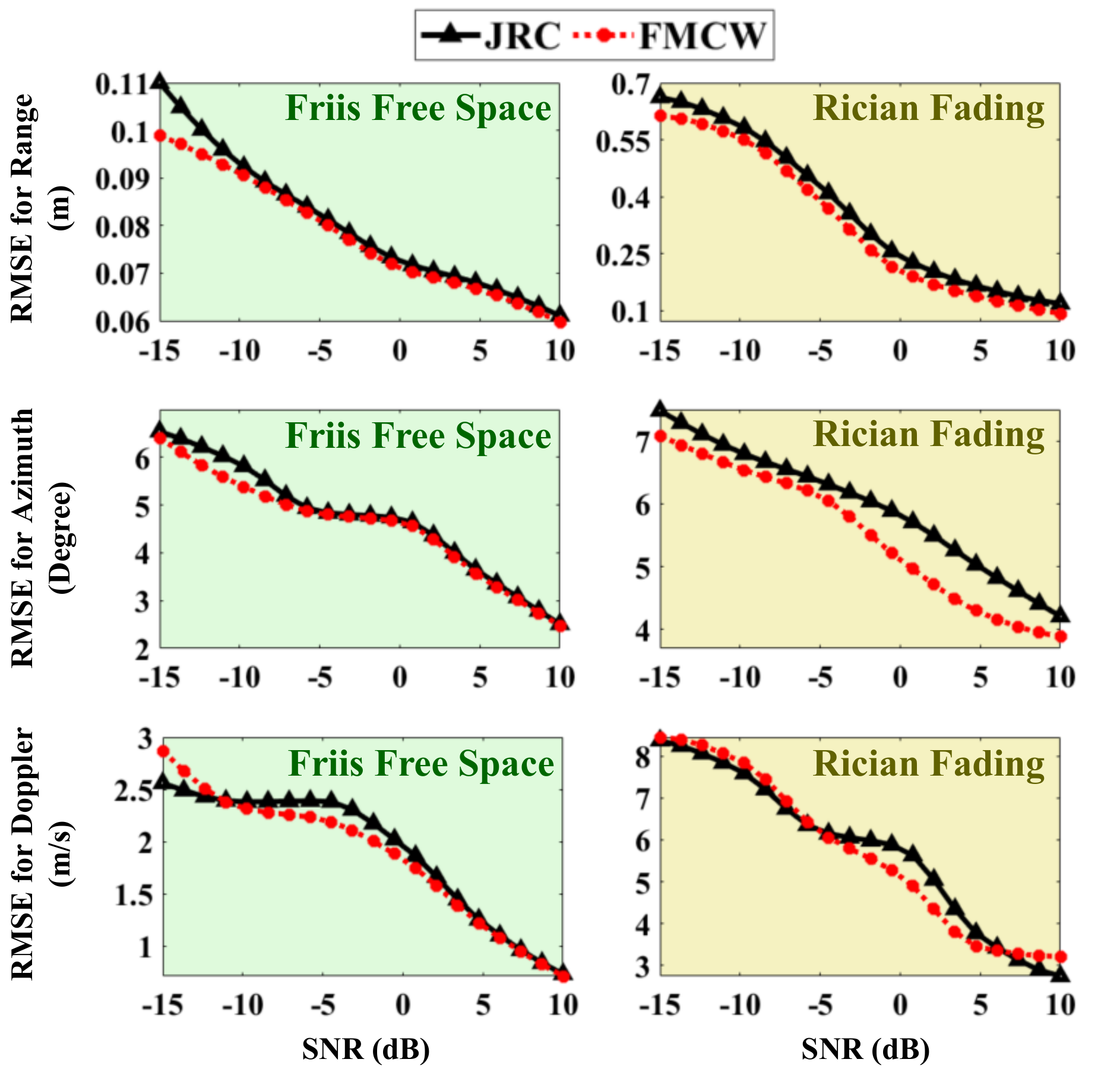}
    \vspace{-6mm}
    \caption{ RMSE results for range, azimuth and Doppler velocity estimates of single pedestrian (extended target) in free space and Rician propagation conditions with JRC and FMCW radar.}
    \label{fig:rmse_human}
    \vspace{-4mm}
\end{figure}

\begin{figure}
\centering
    \includegraphics[scale=0.475]{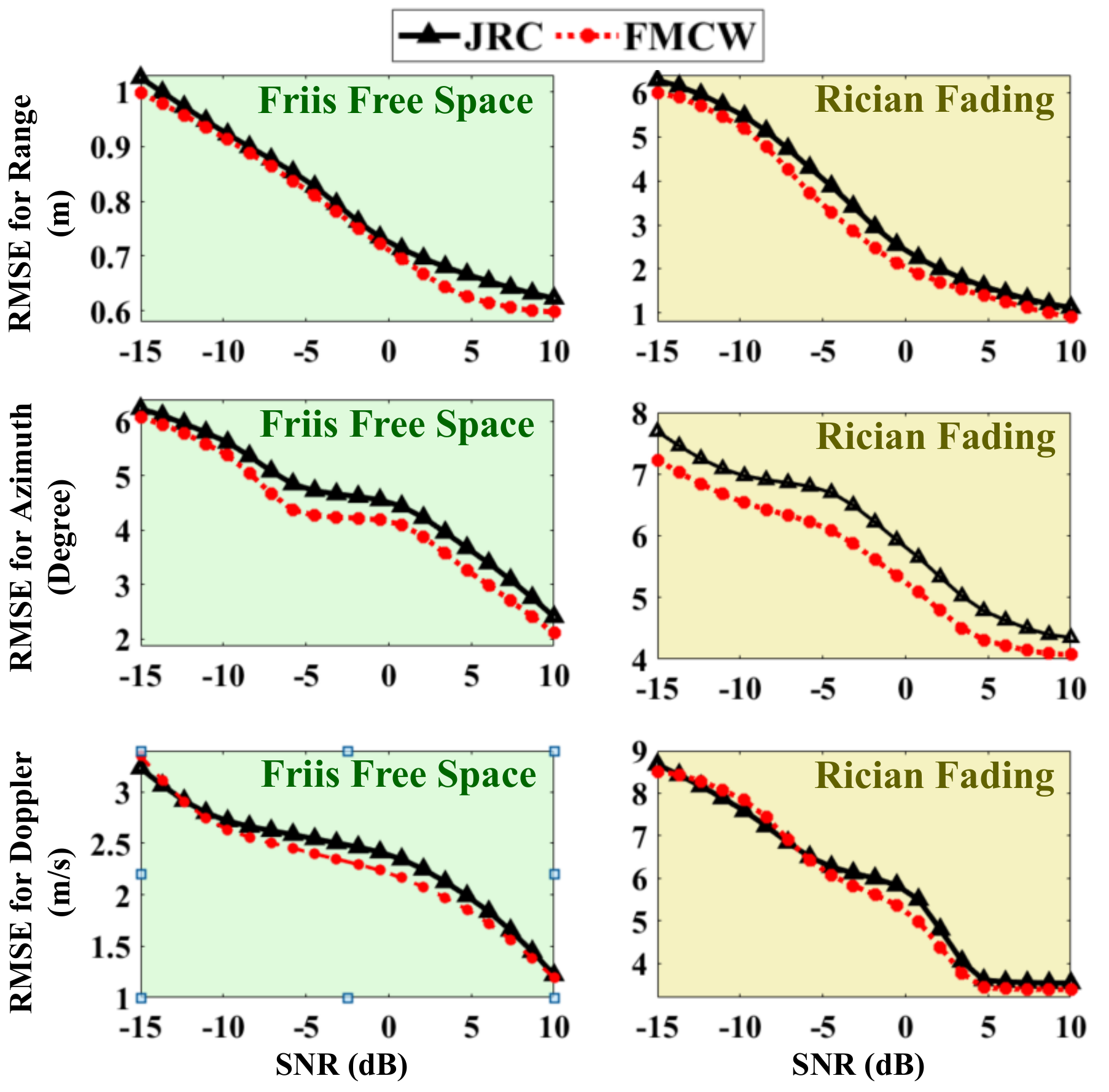}
    \vspace{-4mm}
     \caption{ RMSE results for range, azimuth and Doppler velocity estimates of a single mid-size car (extended target) in free space and Rician propagation conditions with JRC and FMCW radar.}
    \label{fig:rmse_car}
    \vspace{-4mm}
\end{figure}

Next, we repeat the experiment for the two extended dynamic targets - the pedestrian and the midsize car - discussed in Section.\ref{sec:simsetup}. The two targets follow the trajectories shown in Fig.\ref{fig:sim_set}a and b. The ground truth value considered for the calculation of RMSE is the root node point of the multiple-point scatterers on the target. The range, Doppler, and azimuth estimates are reported for free space and Rician propagation conditions using both the JRC waveform and the FMCW in Fig.\ref{fig:rmse_human} and Fig.\ref{fig:rmse_car}. 
The RMSE for a pedestrian deteriorates with lower SNR values under both free space and Rician fading conditions. The pedestrian has a narrower spatial extent along the range dimension when compared to the azimuth width. Additionally, the antenna aperture at the BS is fairly narrow. Due to these reasons, the RMSE error for azimuth is high when compared to that of the range. The  RMSE for Doppler velocity is sufficient to distinguish a moving pedestrian from static clutter. All three estimates deteriorate with the Rician propagation conditions. However, in all of the cases, the results for the JRC waveform are comparable to that of the FMCW radar. In the case of the mid-size car, the RMSE error for range, azimuth, and Doppler are slightly greater than that of the pedestrian since the target is spatially larger and hence the ground truth (at the center of the car) may deviate from the positions of the point scatterers on the chassis of the car that are identified by the radar. 
In this case, the error from the JRC is slightly higher than that of the FMCW waveform (approximately 1 degree in azimuth). Also, the results from the Rician deteriorate slightly with respect to the free space conditions. 
\subsection{Communication Metrics}
\begin{figure}[!t]
\centering
    \includegraphics[scale=0.38]{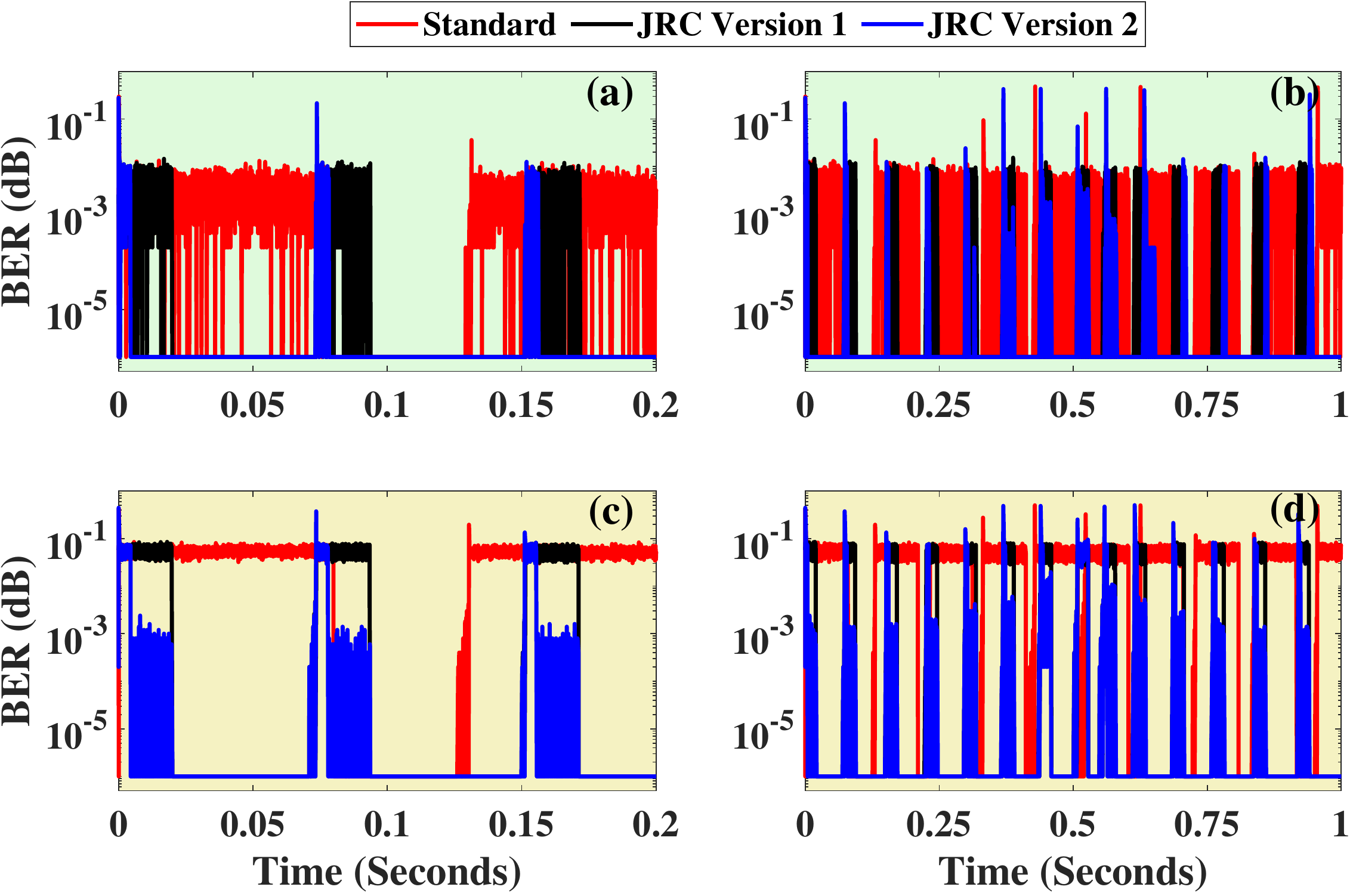}
    \vspace{-4mm}
      \caption{Time-varying BER for Tangential trajectory-1 in free space conditions for (a) short duration of 0.2s, (b) long duration of 1s; and Rician space conditions for (c) short duration of 0.2s, (d) long duration of 1s.}
    \label{fig:comm_results_tra1}
    \vspace{-4mm}
\end{figure}

\begin{figure}[!t]
\centering
    \includegraphics[scale=0.38]{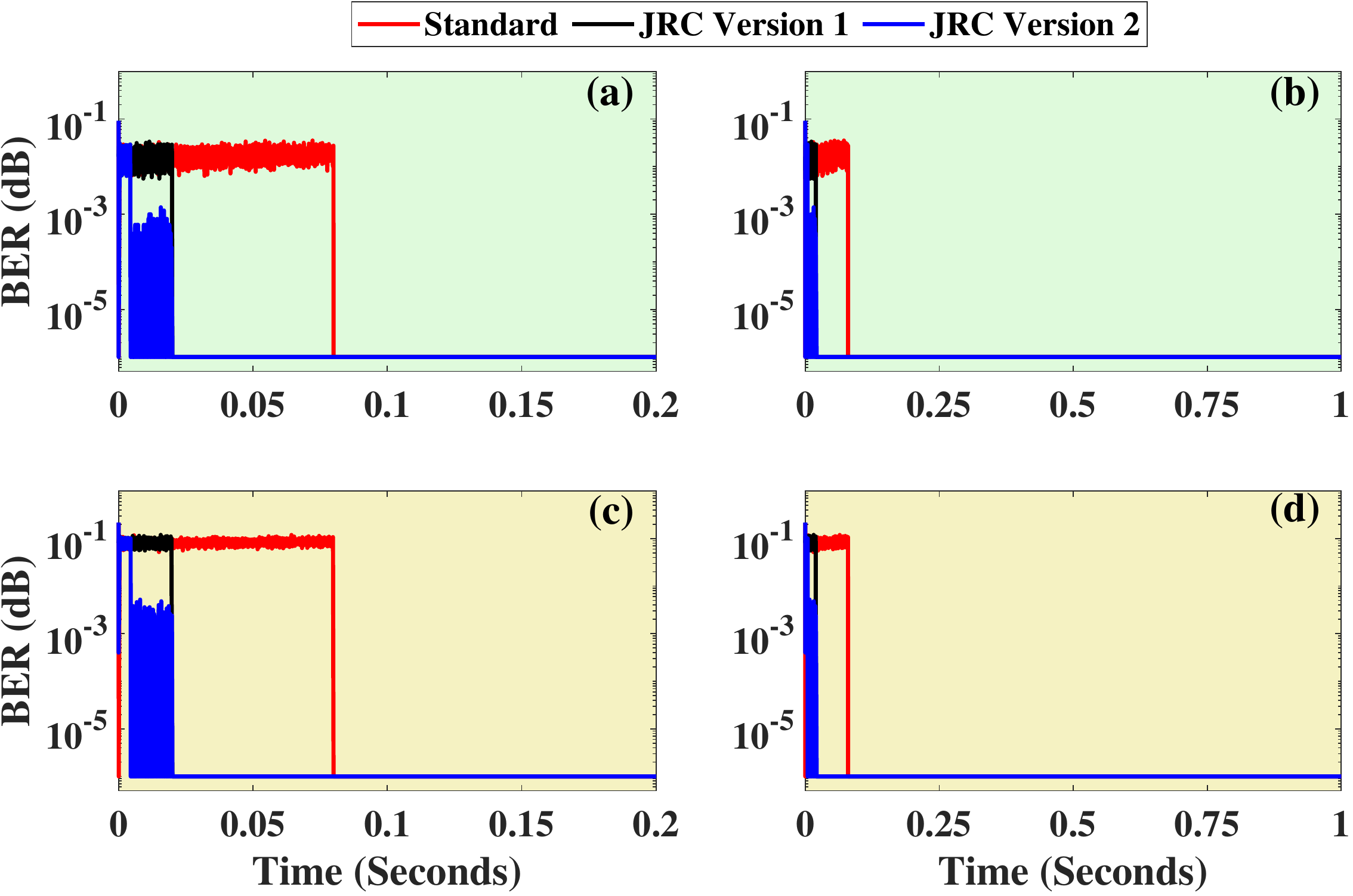}
    \vspace{-4mm}
  \caption{Time-varying BER for radial trajectory-2 in free space conditions for (a) short duration of 0.2s, (b) long duration of 1s; and Rician space conditions for (c) short duration of 0.2s, (d) Long duration of 1s.}
    \label{fig:comm_results_tra2}
    \vspace{-2mm}
\end{figure}

In this section, we report the communication metric - the bit error rate (BER) and throughput -  that are calculated on $\hat{\mathbf{y}}_{DL}$ at MU in Fig.~\ref{fig:comm_results_tra1} and Fig.~\ref{fig:comm_results_tra2}. The BER is the number of error bits out of the total transmitted data bits and is computed for the two different trajectories of the MU and for both free space and Rician channel conditions. The communication receiver is modeled with a system noise temperature of 290K which gives rise to a mean noise floor of -71.7dBm for a receiver bandwidth of 2.64GHz. As the MU moves along the trajectory, the SNR changes based on the distance between the MU and the BS. We plot the BER over the entire duration of each trajectory. 

We first discuss the results for the tangential line trajectory-1 under free space conditions, which are displayed in  Fig.\ref{fig:comm_results_tra1}(a) and (b). 
During \emph{Stage-1}, we observe that the BER of standard and both versions of JRC are high and then fall in Fig.\ref{fig:comm_results_tra1}(a). This corresponds to the poor BER during \emph{Stage-1} because of the poor link metrics arising from the low gain quasi-omnidirectional beams at the BS and MU. This stage is followed by low BER during \emph{Stage-2} with directional links. The \emph{Stage-1} duration for standard, JRC-\emph{Version-1} and JRC-\emph{Version-2} are approximately 0.08s, 0.03s and 0.02s respectively. Hence, the BER is poor for the standard for a longer duration than the JRC versions. 
Interestingly, the BER for \emph{Version-1} falls within the same time as \emph{Version-2} even though theoretically the former takes longer to complete the beam alignment. This is because the BS beam is aligned within this interval for both versions and the high gain from the BS's ULA contributes towards lowering the BER even when the beam alignment is not completed at the MU for the first version. Thus, for free space conditions, we realize an improvement in the beam alignment by a factor of four for both the JRC versions when compared to the standard. Due to the nature of the trajectory, the mobile target crosses a narrow beam within a short duration of time and frequent beam alignment is required as shown in Fig.\ref{fig:sim_set}a. When the target is outside of the beam, we observe a deterioration of the BER due to a fall in the SNR in \emph{Stage-2}. This can be observed, for example, between 0.12s and 0.2s in Fig.\ref{fig:comm_results_tra1}(b) for standard, and between 0.15s to 0.17s for the JRC versions.
Since the standard protocol enables the selection of one of the pre-determined beams while the JRC versions enable directional beams at the recent most angular position of the target, the latter results in better SNR and less frequent beam alignment. The performances of both the standard and the JRC versions deteriorate in Rician conditions resulting in higher BER (as observed in Fig.\ref{fig:comm_results_tra1}(c) and (d)) due to strong interference from multipath. Again, we observe that the beam alignment is more rapid for the JRC versions than standard in Fig.\ref{fig:comm_results_tra1}c which lowers the overall BER by a factor varying from 8 to 32\%.

Next, we discuss the results for the radial trajectory-2 in free space conditions which are displayed in Fig.\ref{fig:comm_results_tra2}(a) and (b). Again, the beam alignment duration for the two versions of JRC is much lower than that of the standard in Fig.\ref{fig:comm_results_tra2}a. The faster beam alignment (again by a factor of four) at the MU for \emph{Version-2} enables slightly superior performance as evidenced by the results. From Fig.\ref{fig:comm_results_tra2}(b), we observe that this trajectory requires less frequent beam alignment between the MU and BS (Fig.\ref{fig:sim_set}b). Once the best beams are identified, the MU remains within the beam for the remaining duration of the motion. The results for the Rician conditions, shown in Fig.\ref{fig:comm_results_tra2} (c) and (d), are much poorer than the free space conditions due to the strong interference from the multipath.
\begin{figure}[!t]
\centering
    \includegraphics[scale=0.5]{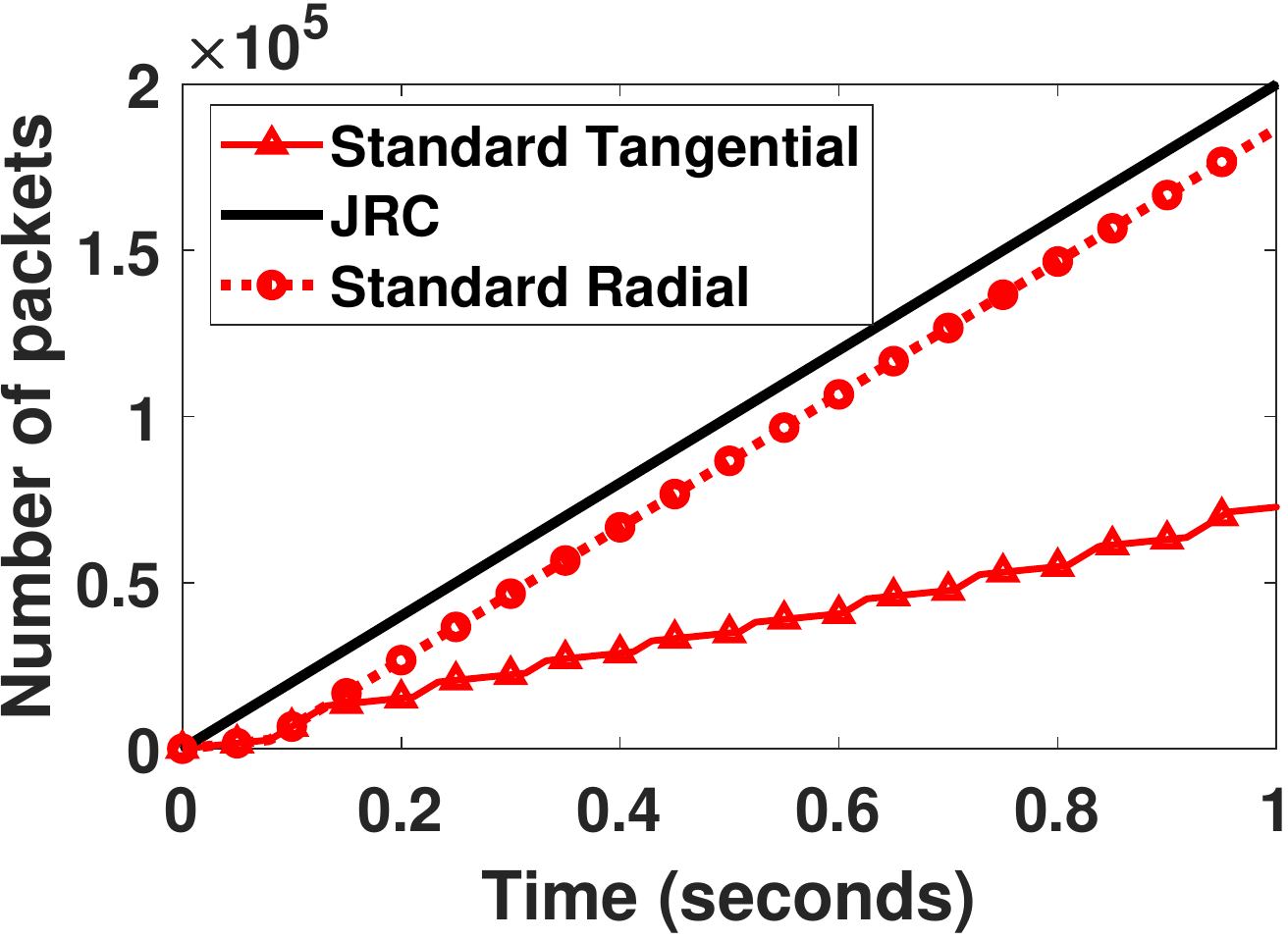}
    \vspace{-3mm}
      \caption{Comparison of number of packets communicated over 1 second duration.}
    \label{fig:comm_packets}
    \vspace{-4mm}
\end{figure}
Next, we compare the number of packets transmitted during the time interval. As shown in Fig.~\ref{fig:comm_packets}, JRC versions allow a higher number of packet transmissions due to smaller packet sizes and fewer beam alignments compared to the standard. The overall communication throughput is calculated based on
\begin{equation}
Throughput = \left(1 - \frac{\sum_{i=1}^{N_p}BER_{i}}{N_p}\right).\frac{(D)(N_p)}{T_d}
\label{eq:Throughput}
\end{equation}
where, $N_p$ is the number of packets transmitted, $T_{d}$ is the total duration and $D$ is the number of data bits transmitted in each packet which is kept fixed.
The overall throughput in Gigabits per second (Gbps) for standard, JRC \emph{Version-1} and JRC \emph{Version-2} are presented in Table.\ref{tab:throughput}. 

\begin{table*}[htbp]
\centering
\caption{Throughput Comparison Of Different Architectures}
\renewcommand{\arraystretch}{1.1}
\begin{tabular}{c|ccc|ccc}
\hline
 &\multicolumn{3}{c|}{Frii's free space} &\multicolumn{3}{c}{Rician}
 \\\hline \hline
Values in Gbps & Standard & JRC \emph{Version-1} & JRC \emph{Version-2}& Standard & JRC \emph{Version-1} & JRC \emph{Version-2}\\\hline
 Trajectory 1 (Translational) & 0.336 & 0.880 & 0.883 & 0.329 & 0.878 & 0.880 \\
 Trajectory 2 (Radial) & 0.840 & 0.883 & 0.883 & 0.839 & 0.881 & 0.883\\ \hline
\end{tabular}
\label{tab:throughput}
\vspace{-3mm}
\end{table*}
The results show that the standard shows a much poorer throughput under both Frii's free space and Rician channel conditions for both trajectories when compared to both versions of the JRC for the first trajectory which required more frequent beam alignments. Even for the second trajectory, the standard has poorer throughput due to the longer beam alignment duration. The throughputs for the two JRC versions are comparable. 
\section{Summary}
\label{sec:Conclusion}
We provide a complete architectural framework of an IEEE 802.11ad-based JRC wireless transceiver for enabling mmW communications between BS and fast-moving users. The radar functionality within the JRC enables the detection of point and extended mobile user targets and provides accurate estimates of their angular positions through range, Doppler, and azimuth processing. This information enables rapid beam alignment of the communication beams at the BS when compared to the lengthy beam training procedure adopted by the standard protocol. A complete end-to-end software prototype implementation on MATLAB is provided with transceiver design details, radar waveform, transmission signal models, and the corresponding signal processing algorithms. For an example case of a system with 32 beams in the BS and 4 beams in the MU, we demonstrate an improvement in the beam alignment timing by a factor of 4 and a significant improvement in the overall communication throughput for the JRC with respect to the standard. The main advantages offered by the JRC are the reduction in the packet lengths and number due to the omission of the beam training fields overhead; and the overall lower BER due to the precise estimate of the MU's angular position by the radar within the BS when compared to the standard's use of pre-determined beams. Further improvements in the JRC timing are anticipated for a higher number of MUs and hardware acceleration of the radar signal processing. This is the focus of our future work. 
\bibliographystyle{ieeetran}
\bibliography{main}

\end{document}